\documentclass[pdflatex,sn-vancouver-num]{sn-jnl}

\usepackage{graphicx}%
\usepackage{multirow}%
\usepackage{amsmath,amssymb,amsfonts}%
\usepackage{amsthm}%
\usepackage{mathrsfs}%
\usepackage[title]{appendix}%
\usepackage[dvipsnames]{xcolor}%
\usepackage{textcomp}%
\usepackage{manyfoot}%
\usepackage{booktabs,makecell}%
\usepackage{algorithm}%
\usepackage{algorithmicx}%
\usepackage{algpseudocode}%
\usepackage{listings}%
\usepackage{tabularx}%
\PassOptionsToPackage{hyphens}{url}\usepackage{hyperref}
\usepackage{xurl}
\usepackage{adjustbox}
\usepackage{subcaption}
\usepackage{array}
\usepackage{float}
\usepackage{caption}

\usepackage{notoccite}

\usepackage{placeins}

\theoremstyle{thmstyleone}%
\theoremstyle{thmstyletwo}%

\theoremstyle{thmstylethree}%

\urldef\YRsbdb\url{https://ssd.jpl.nasa.gov/tools/sbdb_lookup.html#/?sstr=2024%20YR4}

\begin{document}

\title[Article Title]{Space Mission Options for Reconnaissance and Mitigation of Asteroid 2024 YR4}


\author*[1]{\fnm{Brent W.} \sur{Barbee}}\email{brent.w.barbee@nasa.gov} 

\author[2]{\fnm{Matthew A.} \sur{Vavrina}}

\author[3]{\fnm{Rylie} \sur{Bull}}

\author[4]{\fnm{Adrienne} \sur{Rudolph}}

\author[5]{\fnm{Davide} \sur{Farnocchia}}

\author[6]{\fnm{Russell} \sur{TerBeek}}

\author[3]{\fnm{Justin} \sur{Atchison}}

\author[1]{\fnm{Joshua} \sur{Lyzhoft}}

\author[7]{\fnm{Jessie} \sur{Dotson}}

\author[3]{\fnm{Patrick} \sur{King}}

\author[5]{\fnm{Paul W.} \sur{Chodas}}

\author[3]{\fnm{Dawn} \sur{Graninger}}

\author[1]{\fnm{Ronald G.} \sur{Mink}}

\author[8]{\fnm{Kathryn M.} \sur{Kumamoto}}

\author[8]{\fnm{Jason M.} \sur{Pearl}}

\author[8]{\fnm{Mary} \sur{Burkey}}

\author[8]{\fnm{Isaiah} \sur{Santistevan}}

\author[9]{\fnm{Catherine S.} \sur{Plesko}}

\author[9]{\fnm{Wendy K.} \sur{Caldwell}}

\author[9]{\fnm{Megan} \sur{Harwell}}

\affil*[1]{\orgname{NASA/Goddard Space Flight Center}, \orgaddress{\street{8800 Greenbelt Road}, \city{Greenbelt}, \postcode{20771}, \state{MD}, \country{USA}}}

\affil[2]{\orgname{a.i. solutions, Inc.}, \orgaddress{\street{4500 Forbes Boulevard, Suite 300}, \city{Lanham}, \postcode{20706}, \state{MD}, \country{USA}}}

\affil[3]{\orgname{Johns Hopkins University Applied Physics Laboratory}, \orgaddress{\street{11100 Johns Hopkins Road}, \city{Laurel}, \postcode{20723}, \state{MD}, \country{USA}}}

\affil[4]{\orgname{University of Maryland}, \orgaddress{\street{8229 Baltimore Ave}, \city{College Park}, \postcode{20742}, \state{MD}, \country{USA}}}

\affil[5]{\orgdiv{Jet Propulsion Laboratory}, \orgname{California Institute of Technology}, \orgaddress{\street{4800 Oak Grove Dr.}, \city{Pasadena}, \postcode{91109}, \state{CA}, \country{USA}}}

\affil[6]{\orgname{Sandia National Laboratories}, \orgaddress{\street{1515 Eubank Blvd SE}, \city{Albuquerque}, \postcode{87123}, \state{NM}, \country{USA}}}

\affil[7]{\orgname{NASA Ames Research Center, MS 245-6}, \orgaddress{\city{Moffett Field}, \postcode{94035}, \state{CA}, \country{USA}}}

\affil[8]{\orgname{Lawrence Livermore National Laboratory}, \orgaddress{\street{7000 East Avenue}, \city{Livermore}, \postcode{94550}, \state{CA}, \country{USA}}}

\affil[9]{\orgname{Los Alamos National Laboratory}, \orgaddress{\street{P.O. Box 1663}, \city{Los Alamos}, \postcode{87545}, \state{NM}, \country{USA}}}

\abstract{Near-Earth asteroid 2024 YR$_{\text{4}}$ was discovered on 2024-12-27 and its probability of Earth impact in December 2032 peaked at $\sim$3\% on 2025-02-18. Additional observations ruled out Earth impact by 2025-02-23. However, the probability of lunar impact in December 2032 then rose, reaching $\sim$4\% by the end of the apparition in May 2025. James Webb Space Telescope (JWST) observations on 2025-03-26 estimated the asteroid’s diameter at 60 $\pm$ 7 m. Studies of 2024 YR$_{\text{4}}$'s potential lunar impact effects suggest lunar ejecta could increase micrometeoroid debris flux in low Earth orbit up to 1000 times above background levels over just a few days, possibly threatening astronauts and spacecraft. In this work, we present options for space missions to 2024 YR$_{\text{4}}$ that could be utilized if lunar impact is confirmed. We cover flyby \& rendezvous reconnaissance, deflection, and robust disruption of the asteroid. We examine both rapid-response and delayed launch options through 2032. We evaluate chemical and solar electric propulsion, various launch vehicles, optimized deep space maneuvers, and gravity assists. Re-tasking extant spacecraft and using built spacecraft not yet launched are also considered. The best reconnaissance mission options launch in late 2028, leaving only approximately three years for development at the time of this writing in August 2025. Deflection missions were assessed and appear impractical. However, kinetic robust disruption missions are available with launches between April 2030 and April 2032. Nuclear robust disruption missions are also available with launches between late 2029 and late 2031. Finally, even if lunar impact is ruled out there is significant potential utility in deploying a reconnaissance mission to characterize the asteroid.}

\keywords{Planetary Defense, Trajectory Optimization, Asteroid Deflection, Asteroid Disruption}

\maketitle

\section{Introduction}


Near-Earth Asteroid (NEA) 2024 YR$_{\text{4}}$ was discovered on 2024-12-27 by the ATLAS Survey (MPEC 2024-Y140).\footnote{\url{https://www.minorplanetcenter.net/mpec/K24/K24YE0.html}}
\begin{figure}[h]
    \centering
    \includegraphics[width=0.9\textwidth]{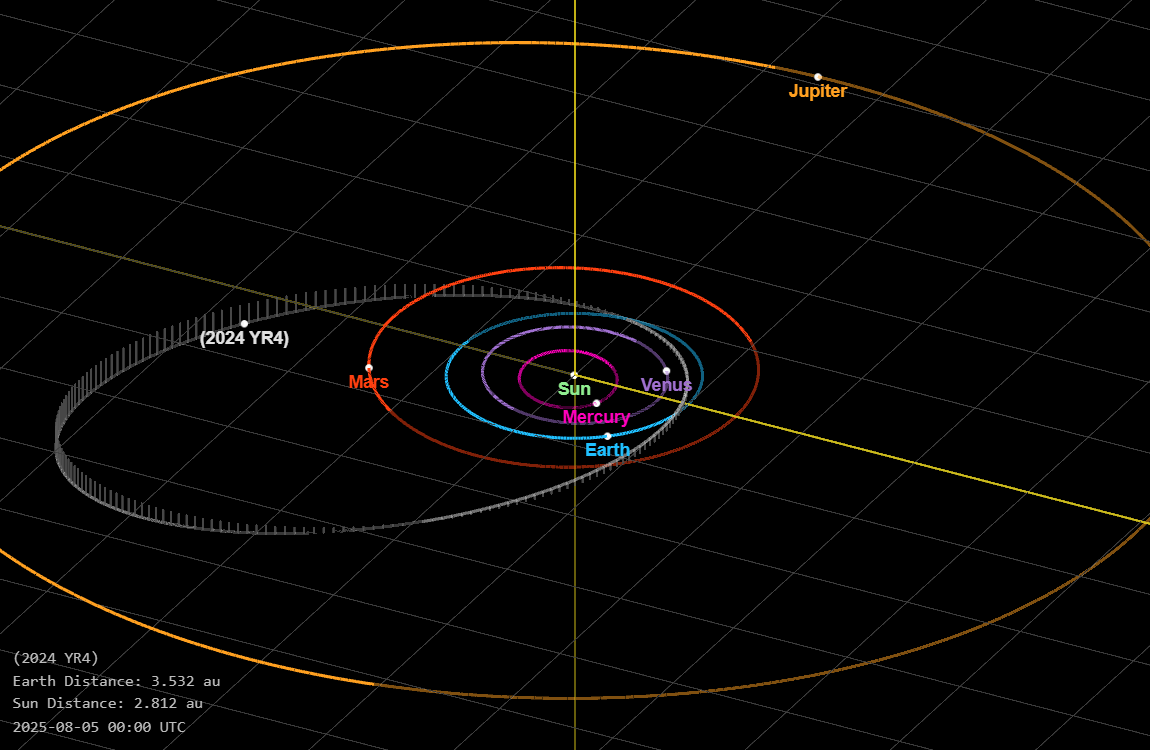}
    \caption{Orbit of asteroid 2024 YR$_{\text{4}}$}
    \label{fig:YR4Orbit}
\end{figure}
Figure~\ref{fig:YR4Orbit} shows a three-dimensional view of the asteroid's orbit, along with the orbits of the inner planets and Jupiter, for reference. This view of the orbit was obtained from the JPL Small-Body Database Lookup\footnote{JPL Small-Body Database Lookup, \url{https://ssd.jpl.nasa.gov/tools/sbdb_lookup.html}, accessed on 2025-08-05.} page for 2024 YR$_{\text{4}}$, and the associated osculating orbital elements are given in Table~\ref{tab:orbelem}.

\begin{table}[h]
\caption{Approximate orbital elements of asteroid 2024 YR$_{\text{4}}$ at epoch 2025-05-05.0 TDB}\label{tab:orbelem}%
\begin{tabular}{cc}
\toprule
Orbital Element & Value \\
\midrule
Semi-major Axis (au) & 2.5159 \\
Eccentricity & 0.6615 \\
Inclination & 3.4082$^{\circ}$ \\
Ascending Node & 271.366$^{\circ}$ \\
Argument of Perihelion & 134.361$^{\circ}$ \\
Mean Anomaly & 40.403$^{\circ}$ \\
Perihelion Distance (au) & 0.8515 \\
Aphelion Distance (au) & 4.1802 \\
\bottomrule
\end{tabular}
\end{table}

Due to its 4 year orbital period, 2024 YR$_{\text{4}}$ will make close approaches to Earth in 2028 and 2032. Initial estimates revealed the possibility of an impact with Earth on 2032-12-22.
As additional tracking data from both the ground and space were collected, the probability initially rose and then peaked at 3\% on 2025-02-18, before falling down to zero in the following days~\cite{Farnocchia_2025}.

However, as the impact probability for the Earth decreased, that of a lunar impact on 2032-12-22 increased and reached 4\% by the end of the discovery apparition in May 2025. Studies of 2024 YR$_{\text{4}}$'s potential lunar impact effects suggest that lunar ejecta could increase micrometeoroid debris flux in low Earth orbit up to 1000 times above background levels over just a few days, possibly threatening astronauts and spacecraft~\cite{Wiegert_2025}.

Due to the significant Earth impact risk, significant physical characterization efforts were undertaken during the discovery apparition. Most notably, the James Webb Space Telescope (JWST) observed 2024 YR$_{\text{4}}$ on 2025-03-08, 2025-03-26, and 2025-05-11. The mid-infrared observations collected on 2025-03-26 constrained the asteroid’s diameter at 60 $\pm$ 7 m and confirmed an S-type taxonomy~\cite{Rivkin_2025}. Additional information about observation efforts is available in~\cite{Bolin_2025}.

An asteroid physical property inference network~\cite{Dotson2024} was used to generate a set of correlated physical property values that are consistent with the JWST results, and our knowledge of the underlying population, while also consisting of physically plausible combinations of properties. Table~\ref{tab:physprop} presents representative examples of the resulting correlated physical property values. The ``Lowest Mass'' and ``Highest Mass'' realizations are the lowest and highest mass physical properties sets, respectively, found in the distribution. However, they are only shown here to provide context because the particular values that comprise them can shift whenever another equivalently possible physical properties distribution is produced. For that reason, we don't regard them as being accurately representative of the lowest and highest mass realizations of the asteroid that should be considered for mission analysis. Instead, we use the low and high ends of the 99.7\% Highest Posterior Density Interval (HPDI) (by mass) for the physical properties distribution, and those two realizations are included in Table~\ref{tab:physprop}. The 25th, 50th, and 75th percentile (\%tile) realizations (by mass) of the asteroid are shown in between the low and high HPDI realizations. For most mission analysis purposes, we will concentrate on three of the realizations in Table~\ref{tab:physprop}: the Low and High 99.7\% HPDI realizations of the asteroid, as bookending cases, and the 50th \%tile realization, which represents the median asteroid.

\begin{table}[htbp]
\centering
\caption{Statistical realizations of 2024 YR$_4$'s physical properties spanning current uncertainties}
\label{tab:physprop}
{\setlength{\tabcolsep}{4pt}
\begin{tabular}{lrrrrrrr}
\toprule
\makecell{\textbf{Physical}\\\textbf{Property}} 
& \thead{Lowest\\Mass}
& \thead{99.7\%\\HPDI Low}
& \thead{25th\\\%tile}
& \thead{50th\\\%tile}
& \thead{75th\\\%tile}
& \thead{99.7\%\\HPDI High}
& \thead{Highest\\Mass} \\
\midrule
Diameter (m)              & 36.64 & 41.9  & 51.11 & 58.42 & 60.51 & 74.7  & 82.16 \\
$\rho_{\text{bulk}}$ (g/cm$^3$) & 1.299 & 1.326 & 2.592 & 2.368 & 2.848 & 3.259 & 3.201 \\
Porosity                   & 0.595 & 0.584 & 0.215 & 0.248 & 0.232 & 0.099 & 0.047 \\
Mass (kg)                       & 3.35E+07 & 5.11E+07 & 1.81E+08 & 2.47E+08 & 3.30E+08 & 7.11E+08 & 9.30E+08 \\
$V_{\text{escape}}$ (cm/s) & 1.56  & 1.80  & 3.08  & 3.36  & 3.82  & 5.04  & 5.50 \\
Abs. Mag                   & 24.015 & 23.827 & 23.965 & 23.877 & 23.987 & 23.883 & 23.995 \\
Albedo                     & 0.326 & 0.296 & 0.175 & 0.146 & 0.123 & 0.089 & 0.066 \\
Tax. Type                  & S     & S     & S     & S     & S     & S     & S     \\
\bottomrule
\end{tabular}
}
\end{table}

This paper is organized as follows. In Section~\ref{s:defl_disr}, we describe the requirements and constraints for the two mitigation options: deflection and robust disruption, including via kinetic impact and nuclear explosive devices. In Section~\ref{s:traj_opts}, we summarize trajectory options for reconnaissance and robust disruption missions, including repurposing existing spacecraft for reconnaissance and developing purpose-built spacecraft for both reconnaissance and robust disruption. Sectiom~\ref{s:ex_missions} presents a set of exemplar mission campaigns. Finally, we summarize the findings and present conclusions in Section~\ref{s:conc}.

\FloatBarrier

\section{Near-Earth Object (NEO) Deflection \& Robust Disruption}\label{s:defl_disr}

Deflection and robust disruption are two strategies for preventing a near-Earth object (NEO) from impacting the Earth. A deflection involves imparting a relatively small $\Delta V$ to the NEO to slightly alter its orbit, while a robust disruption involves applying a much more powerful impulse to the NEO that is designed to blast it into very small and widely scattered fragments. In the following sections we discuss important considerations for deflection and robust disruption, and describe requirements for deflection and robust disruption of 2024 YR$_{\text{4}}$ using Kinetic Impactors (KIs) and Nuclear Explosive Devices (NEDs).

A KI is a spacecraft that impacts the NEO at high velocity in order to change the NEO's momentum, and thereby change the NEO's velocity. In this way, the KI imparts a change in velocity, $\Delta V$, to the NEO. The change in NEO momentum is caused by both the momentum of the spacecraft itself and the momentum carried by any ejecta that may be mobilized out of the impact crater. The additional momentum change from ejecta is represented by a scale factor denoted as $\beta$ and referred to as the momentum enhancement factor (see Eq.~\eqref{eq:KI_DV}). A $\beta$ of 1 means that there is no additional momentum change from the ejecta, a $\beta$ of 2 means the ejecta carries the same amount of momentum as the KI spacecraft effectively doubling the effect, and $\beta > 2$ means the ejecta carries more momentum than the spacecraft. For additional information about KIs, see~\cite{Barbee_2018}. The directionality of the $\Delta V$ vector is of central importance when calculating NEO deflection, and it also matters for the dispersal of post-disruption NEO fragments in directions that will maximally avoid future encounters with the Earth-Moon system. However, for purposes of sizing robust disruption missions we concentrate on the magnitude of the $\Delta V$.

NEDs can, in principle, be detonated against NEOs in standoff, contact, or buried modes. In this work we concentrate on the standoff detonation mode, where the NED is detonated at a selected standoff distance above the asteroid's surface, also referred to as the Height of Burst (HOB). The radiation from the NED detonation will penetrate a thin layer of the asteroid's surface material, rapidly vaporizing that thin layer of material and causing it to blow off of the asteroid towards the zero pressure of vacuum. That vaporized surface material carries significant momentum from the NEO, causing it to recoil in the opposite direction and experience a change in velocity, $\Delta V$, in consequence. For more detailed information about standoff NED detonations versus NEOs, see~\cite{Dearborn_2020}. Analysis of the performance of contact or buried detonation modes is relegated to future work. Note that those modes may require rendezvous with the NEO, or advanced hypervelocity penetrator capabilities for high-speed intercept scenarios. The additional complications of those requirements will have to be carefully weighed against any gains in $\Delta V$ imparted to the NEO for deflection, or improvements in disruption efficacy. By contrast, standoff detonation should be feasible during either rendezvous or high-speed intercept, although rendezvous is generally preferred when available. Additionally, the HOB can be selected during the mission in order to tune the amount of $\Delta V$ imparted to the NEO. Those features provide significant flexibility and greater ability to handle uncertainties.

\subsection{Fragmentation Risks During Deflection} \label{sec21}

The goal when deflecting an NEO is to change its velocity sufficiently to cause it to miss Earth (or the Moon) rather than collide. Additionally, the NEO should either remain intact after the deflection, apart from a relatively small amount of mass that may be ejected by the deflection impulse (e.g., $\sim$ $\leq$ 1\% of the NEO's original mass), or, if the NEO is significantly fragmented, then all or at least a sufficient majority of the fragmented material should have the velocity necessary to miss Earth (or the Moon). Additionally, any resulting NEO debris that passes through the Earth-Moon system should not pose intolerable risks to crew or assets in space or on the lunar surface.

The subsequent motion of fragmented NEO material can be challenging to predict, and an impulse sized for deflection is unlikely to widely scatter the fragments. Therefore, during deflection there is currently a preference for keeping the NEO essentially intact. The current notional heuristic for performing a deflection while avoiding unwanted NEO fragmentation is~\cite{Kumamoto_2025}

\begin{equation}
    \Delta V \leq  10\%~V_{\text{escape}}
    \label{eq:DV_no_frag}
\end{equation}\\

\noindent where $\Delta V$ is that which is imparted to the NEO by an impulsive technique, and $V_{\text{escape}}$ is the NEO's surface escape velocity,

\begin{equation}
    V_{\text{escape}} = \sqrt{\frac{2GM}{R}}
    \label{eq:Vesc}
\end{equation}\\

\noindent and $G$, $M$, and $R$ are the Newtonian constant of gravitation (6.67430 ($\pm$ 0.00015) $\times$ 10$^{-11}$ kg$^{-1}$ m$^3$ s$^{-2}$)\footnote{2018 CODATA recommended values.}, the NEO's mass, and the NEO's body radius (assuming a spherical shape), respectively. When evaluating the required $\Delta V$ for deflecting 2024 YR$_{\text{4}}$, we compare the deflection $\Delta V$ requirements to Eq.~(\ref{eq:DV_no_frag}) to ascertain whether the deflection could be carried out while avoiding unwanted fragmentation. Assessment of whether some degree of NEO fragmentation would be acceptable in cases where the deflection $\Delta V$ exceeds the Eq.~(\ref{eq:DV_no_frag}) criterion is planned for future work.

Although the escape velocity criterion, Eq.~\eqref{eq:DV_no_frag}, is the basis for limiting $\Delta V$ in this work, we are also actively exploring alternative criteria to ensure that fragmentation risks are being mitigated to the greatest extent possible. One such alternative uses the catastrophic disruption threshold, $Q^*_D$, which is extensively used in small body population studies~\cite{Holsapple_2019}. However, the catastrophic threshold alone is insufficient for purposes of specifying the threshold for the onset of unwanted fragmentation because it specifies the specific kinetic energy coupling threshold for a remnant mass precisely half of the original. Scaling laws must therefore be employed to extend this threshold to alternative values more appropriate for specifying the onset of unwanted fragmentation; one such law was proposed by~\cite{Stewart_2009}. Using this prescription would allow one to determine specific energy coupling limits for a given remnant size; notionally, 80\% to 90\% of the original mass are being explored. However, substantial questions remain regarding this methodology, including the applicability of these thresholds to nuclear scenarios; the relative merits of $Q^*_D$-based criteria relative to escape velocity criteria will be evaluated in subsequent work. 

\subsection{Deflection $\Delta V$ Requirements}

The $\Delta V$ that must be imparted to the asteroid in order to deflect it from Earth impact or lunar impact is determined using the B-plane formalism described in~\cite{Farnocchia_2015}. $\Delta V$ applied to an NEO has the most leverage over the NEO's $\zeta$ coordinate in the Earth's B-plane, which corresponds to slowing the NEO down or speeding it up, such that it reaches the Earth (or lunar) encounter earlier or later, respectively. Kinetic Impactors (KIs) usually can only deflect in the negative $\zeta$ direction (slowing the NEO down and therefore reducing its orbital period), while Nuclear Explosive Devices (NEDs) usually can deflect in either the positive or negative $\zeta$ direction (speeding the NEO up or slowing it down, respectively). KIs are generally only able to slow the NEO down because it is very difficult to launch a sufficiently massive KI spacecraft into a \emph{higher} orbit than the NEO, such that the KI spacecraft would be traveling faster than the NEO when impacting it near perihelion. 

The range of $\zeta$ values for 2024 YR$_{\text{4}}$ that correspond to lunar impact are -260500 km to -257872 km. This span is referred to as the impact risk chord, and the mid-point is the center of the chord. The length of the impact risk chord is thus 2628 km. The worst-case situation for a KI is having to deflect the asteroid in the negative $\zeta$ direction all the way across the full chord distance of 2628.3 km. For a standoff NED detonation deflection, the worst-case situation is if the asteroid's lunar impact location is at the mid-point on the chord, requiring the NED to deflect the asteroid a distance of 1314 km in either the negative $\zeta$ or positive $\zeta$ direction. If the asteroid's lunar impact location turns out to be on or near the edge of the impact risk chord in the negative $\zeta$ direction, that would result in a much lower deflection distance requirement, and therefore much lower deflection $\Delta V$ requirement, for KI. This would be true for standoff NED deflection if the asteroid turns out to be impacting on or near the edge of \emph{either} side of the impact risk chord.

\begin{figure}[h]
\centering
\begin{subfigure}{.48\textwidth}
  \centering
  \includegraphics[width=\linewidth]{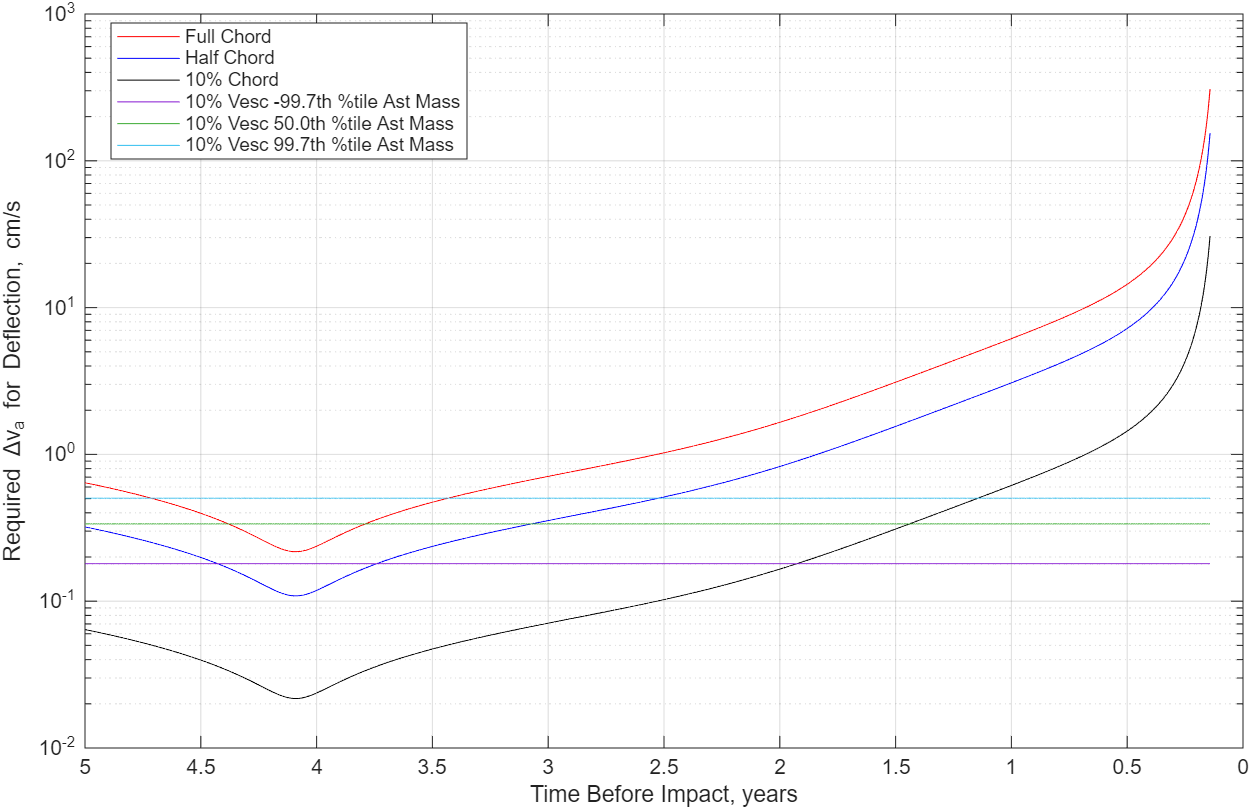}
  \caption{Deflection $\Delta V$ requirements for 2024 YR$_{\text{4}}$ up to $\sim$2 months before lunar impact.}
  \label{fig:Defl_DV_full}
\end{subfigure}%
\hfill
\begin{subfigure}{.48\textwidth}
  \centering
  \includegraphics[width=\linewidth]{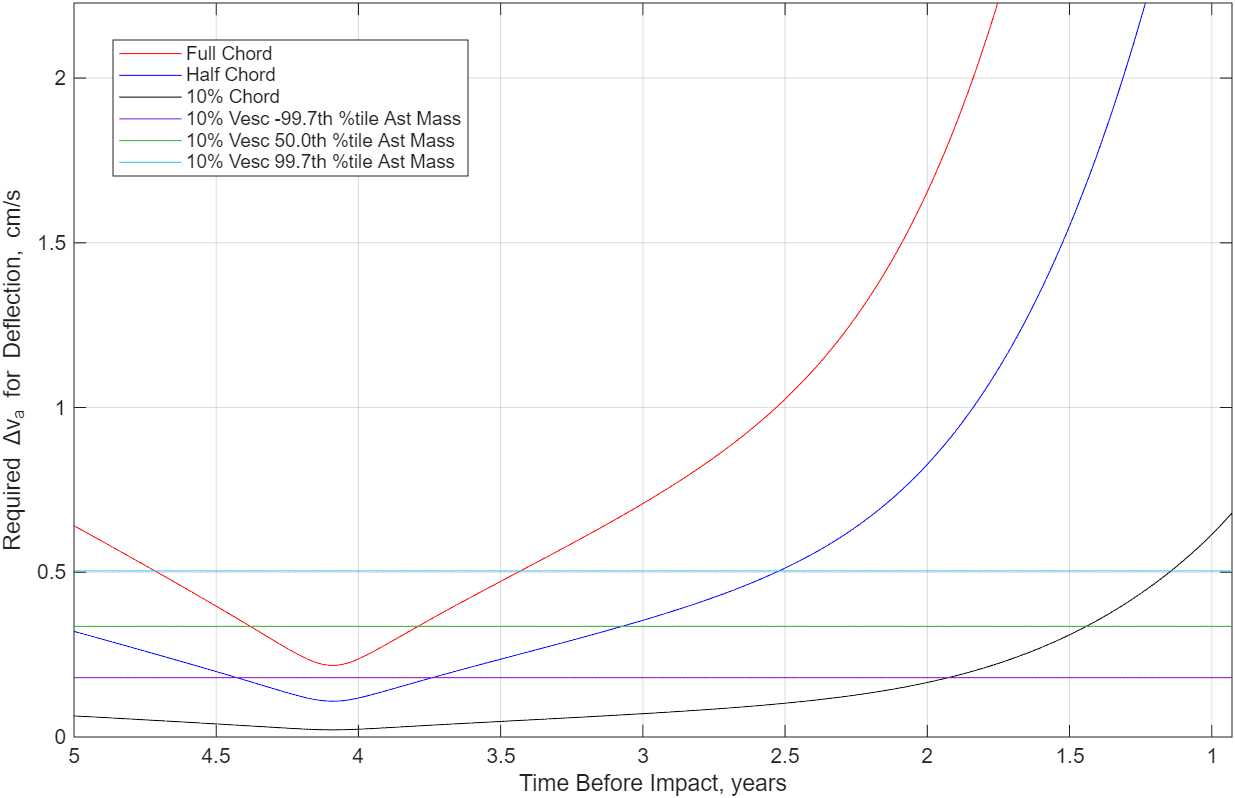}
  \caption{Detail view of Deflection $\Delta V$ requirements relative to 10\% $V_{\text{escape}}$ for 2024 YR$_{\text{4}}$.}
  \label{fig:Defl_DV_zoom}
\end{subfigure}%
\caption{Deflection $\Delta V$ requirements for 2024 YR$_{\text{4}}$ for full, half, and 10\% Impact risk chord length.}
\label{fig:Defl_DV}
\end{figure}

Figure~\ref{fig:Defl_DV} presents the deflection $\Delta V$ requirements as a function of the time (prior to the lunar impact date) at which the deflection is applied to the asteroid. Three curves are shown: the red curve shows the deflection $\Delta V$ required for the full impact risk chord deflection distance, the blue curve is for a deflection distance equal to half of the chord, and the black curve is for a deflection distance equal to 10\% of the chord (to represent the situation where the asteroid's impact location would be near either the positive or negative $\zeta$ edge of the chord).

Figure~\ref{fig:Defl_DV_full} shows that the required deflection $\Delta V$ climbs sharply if applied to the asteroid not long before lunar impact, reaching 31 cm/s for the 10\% chord deflection distance, 155 cm/s for half chord, and 310 cm/s for full chord at $\sim$50 days prior to impact. All of those $\Delta V$ values are far in excess of 10\% of the asteroid's $V_{\text{escape}}$. Figure~\ref{fig:Defl_DV_zoom} includes horizontal lines representing fragmentation thresholds in the form of 10\% $V_{\text{escape}}$ for the 99.7\% HPDI low mass, 50th percentile mass, and 99.7\% HPDI high mass potential realizations of the asteroid. A successful deflection must occur when the $\Delta V$ lies below these lines to avoid undesired fragmentation. The required deflection $\Delta V$ curves cross those fragmentation threshold lines at various different times of deflection.

Figure~\ref{fig:Defl_DV_zoom} shows that the required deflection $\Delta V$ is at a minimum at the time of the asteroid's perihelion in November of 2028: 0.022 cm/s for the 10\% chord deflection distance, 0.11 cm/s for half chord, and 0.22 cm/s for full chord. The full chord deflection $\Delta V$ is never below the 99.7\% HPDI Low Mass fragmentation threshold but remains below the 50th percentile mass fragmentation threshold until the beginning of March 2029. The half chord deflection $\Delta V$ remains below the 99.7\% HPDI Low Mass fragmentation threshold until the end of March 2029, and remains below the 50th percentile mass fragmentation threshold until the end of November 2029. The 10\% chord deflection $\Delta V$ remains below the 99.7\% HPDI Low Mass fragmentation threshold until mid January 2031 and remains below the 50th percentile mass fragmentation threshold until mid July 2031. The available deflection dates could extend somewhat later if the asteroid were to turn out to be at the higher end of the potential mass range, but that is unlikely.

Another aspect of deflection to keep in mind is that one would not want to deflect the asteroid from the vicinity of the Moon all the way towards the Earth and inadvertently cause an Earth impact. The $\zeta$ coordinates in the Earth B-plane corresponding to Earth impact are $\pm$8200 km, and so a very large deflection $\Delta V$ would be needed to deflect the asteroid that far, almost certainly much more than the asteroid could tolerate and remain intact. Nevertheless, it should be kept in mind even though it is an unlikely scenario.

\subsubsection{Gravitational Keyholes}

While a deflection attempt is designed to prevent a specific impact, it is important to consider the secondary risk of keyholes, which are locations on a close encounter's B-plane that correspond to resonant returns: if the asteroid passes through a keyhole, it is injected into an impact trajectory for a later encounter~\cite{Chodas1999}.
While deflecting the asteroid onto a keyhole is a low probability event and a secondary issue compared with the primary impact, higher-fidelity analysis would account for keyholes and attempt to avoid them or at least minimize the chances of hitting one. Table~\ref{tab_keyhole} lists the keyholes for 2024 YR$_{\text{4}}$ corresponding to impact solutions found in the early phases of the discovery apparition. All of the listed keyholes occur far from the Earth and Moon and therefore do not present a design constraint for a mitigation mission, unless a) the asteroid were deflected through the third or fifth keyhole in the table, both of which are relatively near the Moon, b) the asteroid was somehow deflected a much larger distance while remaining intact (very unlikely), or fragments of the asteroid were dispersed widely enough to pass through one or more keyholes. The probability of any of those events occurring is extremely low.

\begin{table}[htbp]
    \centering
    \caption{Keyhole locations on the 2032 Earth B-plane and corresponding impact year for 2024 YR$_{\text{4}}$}
    \setlength{\tabcolsep}{4pt} 
    \begin{tabular}{lc}
        \toprule
        Impact date & Keyhole $\zeta$ (km)\\
        \midrule
        2039-12-23  & -97166\\
        2043-12-23  & -33177\\
        2047-12-22 & -267277\\
        2047-12-22 & -188497\\
        2047-12-22 & -264143\\
        2079-12-22  & -40763\\
        \bottomrule
        \end{tabular}
    \label{tab_keyhole}
\end{table}

\subsection{Robust Disruption Requirements}

A NEO disruption is considered \textit{robust} if the NEO is forcefully broken up into many small fragments no larger than 10 m in size, all of which are scattered with enough velocity to be widely dispersed. Additionally, the robust disruption should be performed far enough in advance of the Earth-Moon system encounter date that any resulting post-disruption NEO debris flux through the Earth-Moon system will be low enough to be deemed tolerable. Ideally, very little, if any, NEO material would interact with the Earth, the Moon, or any astronauts or space assets in the Earth-Moon system.

Our current notional heuristic for robust disruption is given by~\cite{Kumamoto_2025}

\begin{equation}
    \Delta V \geq 10 \times V_{\text{escape}}
    \label{eq:robustD_DV}
\end{equation}\\

\noindent However, similar to the discussion surrounding thresholds for avoiding fragmentation in Section \ref{sec21}, alternatives based on $Q^*_D$ can be formulated. In this case, robust disruption - because fragmentation to a specific size threshold is required - introduces further size-dependence than $Q^*_D$ alone, and larger asteroids become progressively more difficult to robustly disrupt. Nevertheless, similar questions remain about the applicability of the $Q^*_D$ values reported in \cite{Holsapple_2019} and elsewhere to non-kinetic loading, and so we will defer the consideration of these thresholds to future work and utilize Eq.~\eqref{eq:robustD_DV} for these results. 

\subsubsection{Kinetic Robust Disruption}

\begin{figure}[h]
    \centering
    \includegraphics[width=0.7\textwidth]{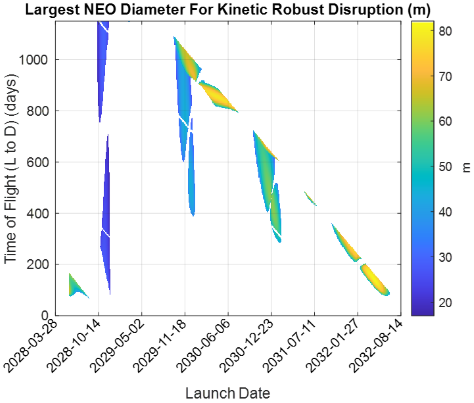}
    \caption{Pork Chop Contour (PCC) plot of maximum 2024 YR$_{\text{4}}$ diameter that is robustly disruptable via KI.}
    \label{fig:KI_disr_PCC}
\end{figure}

To identify feasible and optimal KI spacecraft trajectories for kinetic robust disruption, we begin with Eq.~\eqref{eq:robustD_DV} and Eq.~\eqref{eq:Vesc}. We then substitute the equation for spherical NEO mass, given by  

\begin{equation}
    M = \rho\frac{4}{3}\pi R^3
    \label{eq:NEOmass}
\end{equation}\\

\noindent where $\rho$ is the NEO's bulk density and $R$ is the NEO's spherical radius, into Eq.~\eqref{eq:Vesc}, and then substitute the resulting expression into Eq.~\eqref{eq:robustD_DV}. Next, we utilize the equation for the $\Delta V$ imparted to an NEO via a KI, given by

\begin{equation}
    \Delta V = \frac{\beta m_{\text{KI}} v_{\text{rel}}}{M}
    \label{eq:KI_DV}
\end{equation}\\

\noindent where $\beta$ is the momentum enhancement factor associated with the kinetic impact, representing the momentum carried by ejecta produced by the KI impacting the NEO, $m_{\text{}KI}$ is the mass of the KI spacecraft, and $v_{\text{rel}}$ is speed of the KI spacecraft relative to the NEO at the time of impact. We substitute Eq.~\eqref{eq:NEOmass} into Eq.~\eqref{eq:KI_DV} and then substitute the resulting expression into Eq.~\eqref{eq:robustD_DV}. After algebraic reduction, we obtain an expression for the maximum spherical NEO diameter that can be robustly disrupted kinetically,

\begin{equation}
    D_{\text{max}} = \left[\frac{54 \beta^2 v^2_{\text{rel}} m^2_{\text{KI}}}{G \rho^3 \pi^3 F^2}\right]^{\frac{1}{8}}
    \label{eq:Dmax}
\end{equation}\\

\noindent where $F = \Delta V / V_{\text{escape}} = 10$ (via Eq.~\eqref{eq:robustD_DV}). We then encode Eq.~\eqref{eq:Dmax} into our Lambert grid scan code and perform a grid scan, which reports the maximum robustly disruptable NEO diameter for each combination of launch date of time of flight. The robustly disruptable NEO diameters indicated by this analysis are compared to the range of NEO diameters spanned by the 99.7\% HPDI in Table~\ref{tab:physprop} to assess whether kinetic robust disruption is viable for 2024 YR$_{\text{4}}$. The Pork Chop Contour (PCC) plot showing these grid scan results is presented in Fig.~\ref{fig:KI_disr_PCC}.

To generate the data in Fig.~\ref{fig:KI_disr_PCC}, Eq.~\eqref{eq:Dmax} uses $\beta = \text{2}$ and $\rho$ equal to the largest density value in Table~\ref{tab:physprop}, 3.259 g/cm$^3$ (associated with the 99.7\% HPDI High Mass realization of the asteroid). Note that $\beta = \text{2}$ is based on the DART results~\cite{Cheng_2023} but $\beta$ could be significantly higher than 2, particularly for very high velocity impacts. The largest robustly disruptable diameter for 2024 YR$_{\text{4}}$ yielded by this analysis is approximately 81 m, which is larger than the 74.7 m diameter of the 99.7\% HPDI High Mass realization of the asteroid and is nearly equal to the diameter of the overall Highest Mass realization of the asteroid (which has a slightly lower bulk density associated with it).

Altogether, these results imply that deploying a KI spacecraft with mass and trajectory associated with the highest robustly disruptable diameter values in Fig.~\ref{fig:KI_disr_PCC} should be able to robustly disrupt any physical realization of the asteroid within the current asteroid properties distribution. This would provide some confidence of disruption mission success even without first sending a reconnaissance mission to the asteroid, although we still have a strong preference for reconnoitering the asteroid prior to building and deploying the disruption mission if at all possible, and for at least having a rendezvous observer near the asteroid to provide situational awareness and debris tracking during and after the disruption attempt.

Another important consideration in robust disruption is the time provided for the post-disruption asteroid debris to spread out and disperse, because that will strongly influence the overall level of micrometeoroid debris flux through the Earth-Moon system (which can pose risks to astronauts, spacecraft, and lunar surface assets), the total mass and energy of debris that impacts Earth's atmosphere, and whether that atmosphere-impacting debris includes any sizeable (i.e. several meter size) fragments. Ideally, the robust disruption would be performed far enough in advance of the Earth encounter date for the post-disruption debris to disperse such that subsequent debris passage through the Earth-Moon system is avoided altogether or at least minimized. Previous work~\cite{King2021LateTimeDisruptions} suggests that the fraction of Earth-impacting (or Moon-impacting) NEO mass can be reduced by 2 to 3 orders of magnitude if a robust disruption is performed at least one month prior to the Earth encounter date. The higher end of the reduction in post-disruption impacting NEO mass tends to be associated with more eccentric orbits, which applies to 2024 YR$_{\text{4}}$. The Moon also has a smaller gravitational capture cross-section than the Earth, which should further reduce gravitational focusing effects on post-disruption NEO material. Furthermore, the scenarios studied in~\cite{King2021LateTimeDisruptions} involve larger NEOs than 2024 YR$_{\text{4}}$. Altogether, this suggests that robustly disrupting 2024 YR$_{\text{4}}$ at least one month before lunar encounter would be sufficient to limit post-disruption debris effects in the Earth-Moon system. However, to add some margin, in our current work we assume that the robust disruption, whether kinetic or nuclear, should occur no later than 3 months before the lunar impact date, i.e. no later than 2032-09-22. Further work is necessary to develop specific requirements for how far in advance of Earth encounter a robust disruption should be performed and how the quality of the result should be measured, possibly by comparison of predicted post-disruption debris flux through the Earth-Moon system to the natural background levels, and establishing tolerable levels of transient debris flux above the background.

\subsubsection{Nuclear Robust Disruption}

\begin{figure}[h]
    \centering
    \includegraphics[width=0.6\textwidth]{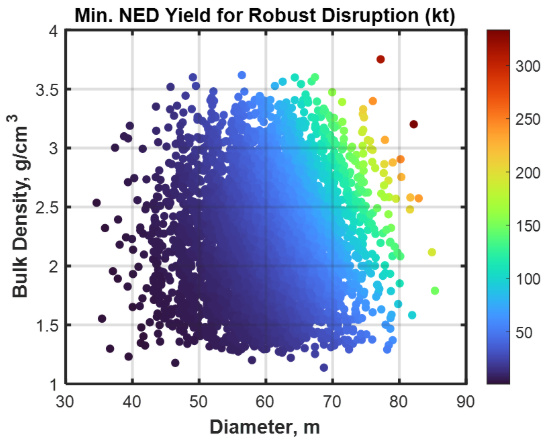}
    \caption{Minimum NED yield required to robustly disrupt the potential realizations of 2024 YR$_{\text{4}}$.}
    \label{fig:NED_yield_disr}
\end{figure}

In this work, we use an approximate analytical model~\cite{Managan_2025} for calculating the $\Delta V$ imparted to an NEO via an NED, as a function of NED yield, Height of Burst (HOB) above the NEO's surface, NEO body radius, NEO bulk porosity, and NEO bulk density. We use that model with Eq.~\eqref{eq:robustD_DV}, Eq.~\eqref{eq:Vesc}, and the distribution of physical properties from which the particular potential realizations of 2024 YR$_{\text{4}}$ given in Table~\ref{tab:physprop} are drawn to calculate the minimum NED yield (detonated at optimal HOB that maximizes $\Delta V$ imparted to the NEO) required to robustly disrupt all combinations of NEO diameter, density, and porosity.\footnote{The analytical model for $\Delta V$ imparted to an NEO by a standoff NED detonation was intended for estimating achieved $\Delta V$ and HOB for a given NED yield. It was not intended to be used for NEO disruption modeling, so our use of it in that capacity should be regarded as only a rough approximation. NEO disruption via NED detonation must be assessed using appropriate techniques, including radiation transport codes and hydrocodes.} The results of this analysis are summarized in Figure~\ref{fig:NED_yield_disr}.

\begin{figure}[h]
\centering
\begin{subfigure}{.45\textwidth}
  \centering
  \includegraphics[width=\linewidth]{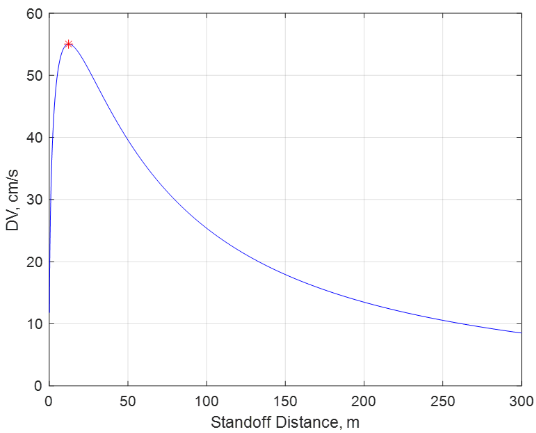}
  \caption{334 kt NED yield.}
  \label{fig:334ktHOB}
\end{subfigure}%
\hfill
\begin{subfigure}{.45\textwidth}
  \centering
  \includegraphics[width=\linewidth]{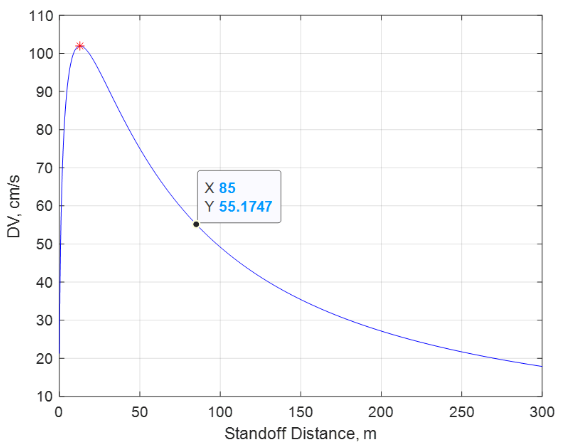}
  \caption{1 Mt NED yield.}
  \label{fig:1MtHOB}
\end{subfigure}%
\caption{$\Delta V$ imparted to Highest Mass realization of 2024 YR$_{\text{4}}$ as a function of standoff distance (HOB) for two NED yields.}
\label{fig:NEDHOBs}
\end{figure}

According to Figure~\ref{fig:NED_yield_disr}, the largest required NED yield for robustly disrupting 2024 YR$_{\text{4}}$ is $\sim$334 kt, corresponding to the Highest Mass realization of the asteroid. For the 99.7\% High Mass realization of the asteroid, the required NED yield for robust disruption is $\sim$204 kt. However, it must also be noted that those yields correspond to detonating the NED at the optimal HOB that maximizes $\Delta V$ imparted to the NEO. As shown in Figure~\ref{fig:334ktHOB}, the HOB for the 334 kt NED detonation would be only ~12 m, which would provide essentially no margin for error during the timing of the detonation. That would not be operationally robust, especially if detonating the NED during a high-speed intercept of the asteroid rather than after rendezvous.

Figure~\ref{fig:1MtHOB} shows that with a 1 Mt NED, the necessary $\Delta V$ of $\sim$55 cm/s for robust disruption of the Highest Mass realization of 2024 YR$_{\text{4}}$ is achieved at a HOB of 85 m. That provides more margin for the detonation operations, and, for smaller realizations of the asteroid, even higher HOBs are acceptable and provide even more margin. 1 Mt is therefore selected as a useful NED yield that could be deployed for robust disruption of the asteroid even without a prior reconnaissance mission. However, that would require designing the spacecraft to be effective at the smallest HOB, required for the largest potential realization of the asteroid. If a recon mission reveals that the asteroid is smaller than the largest possible size, that would mean that a higher HOB would be acceptable, potentially easing the mission and spacecraft design requirements.

Additionally, we note once again that we have a strong preference for a rendezvoused observer spacecraft to monitor the asteroid before, during, and after the detonation. If a rendezvous mission is capable of delivering a deployable NED to the asteroid, then that NED delivery spacecraft could serve dual-purpose as the observer spacecraft. This could be accomplished by, for example, moving the spacecraft to the opposite side of the asteroid, and at an appropriate distance, after the deployment of the free-flying NED, to use the asteroid itself as a shield during the detonation. However, because rendezvous trajectory options are very limited for 2024 YR4, we will be emphasizing mission profiles involving detonating the NED during a high-speed intercept of the asteroid.

Table~\ref{tab:NEDtimes} summarizes the maximum effective HOB for robust disruption using a 1 Mt NED against the potential realizations of the asteroid listed in Table~\ref{tab:physprop}. These provide reference values against which the time required for detonation operations, including arming, firing, and fuzing, can be compared to ensure the combination of intercept speed and HOB utilized for the mission provide adequate time for detonation to occur reliably. Those assessments are future work. Additionally, the NED radar system must be able to fuze at the HOB chosen for the mission.

\begin{table}[h]
\caption{Time available to detonate a 1 Mt NED, in milliseconds, for robust disruption during high-speed intercept, as a function of intercept speed and HOB for several potential realizations of 2024 YR$_{\text{4}}$ that span current uncertainties in the asteroid's physical properties.}\label{tab:NEDtimes}%
\begin{tabular}{lccccccccc}
\toprule
Asteroid    & Max. & 3 & 5 & 10 & 15 & 20 & 25 & 30 & 35 \\
Realization & HOB & km/s & km/s & km/s & km/s & km/s & km/s & km/s & km/s \\
\midrule
Highest Mass & 85 & 28 & 17 & 9 & 6 & 4 & 3 & 3 & 2 \\
99.7\% HPDI High & 113 & 38 & 23 & 11 & 8 & 6 & 5 & 4 & 3 \\
75th Percentile & 217 & 72 & 43 & 22 & 14 & 11 & 9 & 7 & 6 \\
50th Percentile & 277 & 92 & 55 & 28 & 18 & 14 & 11 & 9 & 8 \\
25th Percentile & 329 & 110 & 66 & 33 & 22 & 16 & 13 & 11 & 9 \\
99.7\% HPDI Low & 740 & 247 & 148 & 74 & 49 & 37 & 30 & 25 & 21 \\
Lowest Mass & 888 & 296 & 178 & 89 & 59 & 44 & 36 & 30 & 25 \\
\bottomrule
\end{tabular}
\end{table}

\subsubsection{NED Radar Fuze Modeling for 2024 YR$_{\text{4}}$} \label{ss:radar_fuze}

The radar fuze is a critical component of an NED-based deflection or disruption mission. In order to determine the suitability of radar for any given planetary defense mission, we run Sandia National Laboratory's (SNL's) radar simulation code, Probability of Failure due to insufficient Signal (PFS), which determines the PFS for the given scenario. The code calculates the received power per echo pulse and integrates those pulses over time. If the time-integrated power surpasses the threshold signal-to-noise ratio, then the object is successfully detected. There is also a Monte Carlo element to the simulation: small variations are taken with respect to the radar position, velocity, and timing to see how sensitive the final probability is to such departures from design parameters.

The SNL PFS code was initially built with a ``flat-earth'' geometry. A given mission is uniquely specified by the radar's initial height, velocity, and declination angle with respect to the horizontal. We have rebuilt PFS to account for arbitrary input geometries. This is an important correction, because a completely flat geometry will overestimate the actual radar returns with respect to a surface that curves away from the radar. In general, radar terrain backscattering tends to be an exponentially decaying function of incident angle~\cite{Ulaby_2019}.

In order to study the case of asteroid 2024 YR$_{\text{4}}$, we make a few modeling assumptions:

\begin{enumerate}

    \item The antenna pattern is isotropic.
    \item We model the asteroid as an ellipsoid of dimensions 87 m $\times$ 87 m $\times$ 29 m. This is based on computing an ellipsoid with the possible axis ratios of 3:3:1 reported in~\cite{Bolin_2025} while preserving the volume of the 60 m nominal asteroid reported in~\cite{Rivkin_2025}.
    \item The surface of the asteroid is carved up into triangular facets of edge length on the order of tens of wavelengths. The axioms of radar terrain scattering theory only hold when the scattering object is electrically large.
    \item Each facet is modeled as a point scatterer with an amplitude that depends on the angle between its normal vector and the incoming EM. All facets within the beam ``footprint'' contribute to the received power.
    
\end{enumerate}

\begin{figure}[h]
    \centering
    \includegraphics[width=.8\textwidth]{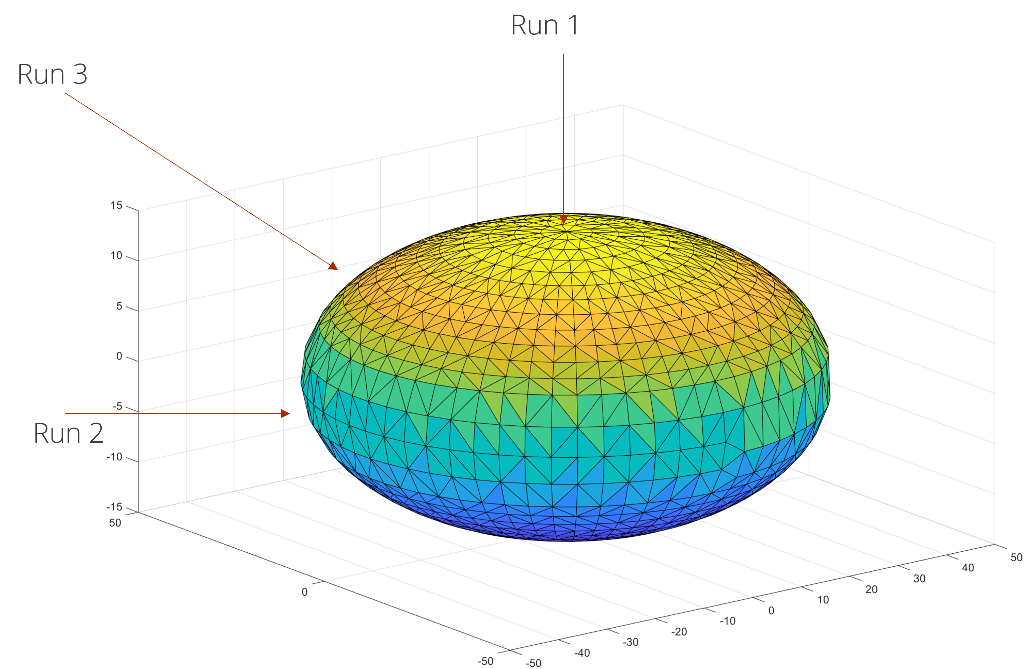}
    \captionof{figure}{Three geometries considered in radar fuze modeling for 2024 YR$_{\text{4}}$, in terms of spacecraft velocity vectors relative to asteroid during terminal approach.}
    \label{fig:radar}
\end{figure}

The geometries used for the various radar model runs are depicted in Figure~\ref{fig:radar}. In Run 1, the NED-carrying spacecraft approaches the asteroid along a line normal to the asteroid's widest face. In Run 2, the NED is approaching the asteroid normal to its narrowest face, i.e. edge on. In Run 3, the NED is approaching along a line inclined 45$^{\circ}$ to the plane of the asteroid's widest face, i.e., halfway between Runs 1 and 2. The results of the model runs are shown in Table~\ref{tab:radar}. Approach speed was varied from 2 km/s to 35 km/s for each of the three relative velocity vector directions relative to the asteroid's shape shown in Figure~\ref{fig:radar}. Table~\ref{tab:radar} reports the PFS\footnote{Probabilities less than $1 \times 10^{-5}$ are rounded up to that number.} for each combination of approach direction and speed. Intolerably high probabilities of failure are color-coded red. The remaining probability values are considered tolerable.

\begin{table}[h]
\caption{Probability of NED radar fuzing failure due to insufficient signal for 2024 YR$_{\text{4}}$, as a function of approach angle and speed.}\label{tab:radar}
\begin{tabular}{ccccccccc}
\toprule
 & 2 km/s & 5 km/s & 10 km/s & 15 km/s & 20 km/s & 25 km/s & 30 km/s & 35 km/s \\
\midrule
Run 1 & 1e-5 & 1e-5 & 3.308e-5 & 7.88e-4 & 3.6e-3 & \textcolor{red}{0.8094} & \textcolor{red}{1} & \textcolor{red}{1} \\
Run 2 & 1e-5 & 1e-5 & 6.15e-5  & 1.4e-3  & \textcolor{red}{0.714}  & \textcolor{red}{1}      & \textcolor{red}{1} & \textcolor{red}{1} \\
Run 3 & 1e-5 & 1e-5 & 4.07e-5  & 1.0e-3  & \textcolor{red}{0.955}  & \textcolor{red}{1}      & \textcolor{red}{1} & \textcolor{red}{1} \\
\bottomrule
\end{tabular}
\end{table}

In this scenario, subject to the model assumptions and timing configuration of the radar (which is admittedly arbitrary), the radar fuzed successfully for all approach velocities up to and including 15 km/s. One run succeeded at 20 km/s; it is expected that the simulation will yield success for some approach trajectories but not others, and the trajectory that fuzed successfully at 20 km/s is Run 1, which had maximum asteroid surface area in the plane normal to the spacecraft's approach direction. For the runs that failed, it is possible to tweak the Doppler filter in order to ensure success at higher approach speeds at the cost of reducing signal-to-noise ratio for all approach speeds. The fuzing height above the asteroid surface, which bounds the HOB from above, was approximately 100--104 m for all successful runs. These results illustrate the importance of performing radar simulations ahead of time to identify performance problems that could impact mission design. With enough advance notice, these results can be used to modify the hardware design and ensure a successful nuclear deflection or disruption of the asteroid.

\section{Mission Trajectory Options}\label{sec_trajOptions}\label{s:traj_opts}


\subsection{Use of Extant Spacecraft}

In some cases, it may be possible to repurpose or retask an extant spacecraft for rapid reconnaissance. This can enable in-space observations faster than a purpose-built spacecraft could arrive at the asteroid at the cost of the original mission. Bull et al.~\cite{Bull_2025} detail the pros and cons of this approach, and the considerations that are necessary beyond trajectory reachability. We assessed the following extant spacecraft for their potential to reconnoiter 2024 YR$_{\text{4}}$: Janus, Lucy, OSIRIS-APEX, and Psyche. We determined that utilizing Lucy for 2024 YR$_{\text{4}}$ would most likely be infeasible given high $\Delta V$ demand and propellant limitations. Possibilities for utilizing the other spacecraft are discussed in turn.

\subsubsection{Janus}

The Janus spacecraft are two small spacecraft originally designed for flybys of binary asteroid systems, built under NASA's SIMPLEx program~\cite{Shoer_2022}. They were originally scheduled to launch as a rideshare with the Psyche spacecraft, but were left without a launch opportunity when Psyche's launch was significantly delayed \cite{scheeres2024}. The spacecraft were then partially disassembled and placed into storage. While a full engineering analysis of the mission trajectory performance envelope for which Janus would function satisfactorily is beyond the scope of our current work, we do consider two Janus performance limitations when assessing potential 2024 YR$_{\text{4}}$ flyby trajectories for which Janus could potentially work: 1) Sun distance must remain between 1 and 1.62 au; 2) Earth distance must remain less than 2.4 au \cite{Shoer_2022}. 

The $\sim$4-year orbit period and $\sim$4 au aphelion of 2024 YR$_{\text{4}}$'s orbit lead us to expect that in order for a spacecraft to launch from Earth and encounter the asteroid before it has passed beyond 1.62 au from the Sun, a relatively high-energy launch must be performed sometime during the months leading up to the asteroid's perihelion, or during the time period surrounding perihelion itself.

\begin{figure}[h]
\centering
\begin{subfigure}{.45\textwidth}
  \centering
  \includegraphics[width=\linewidth]{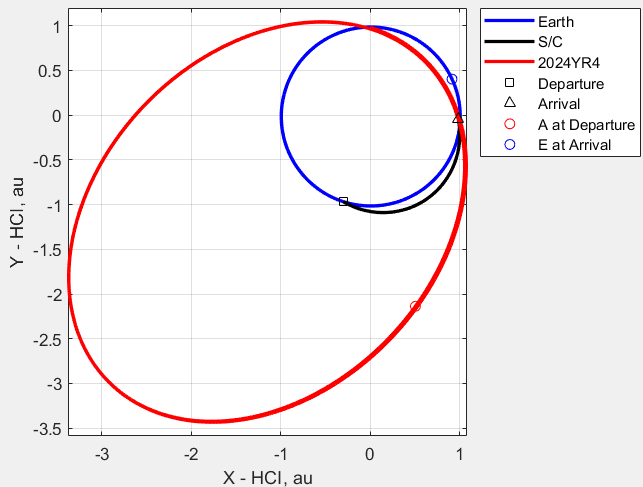}
  \caption{Exemplar 2024 YR$_{\text{4}}$ flyby reconnaissance trajectory using Janus with a June 2028 launch}
  \label{fig:janusJun2028}
\end{subfigure}%
\hfill
\begin{subfigure}{.45\textwidth}
  \centering
  \includegraphics[width=\linewidth]{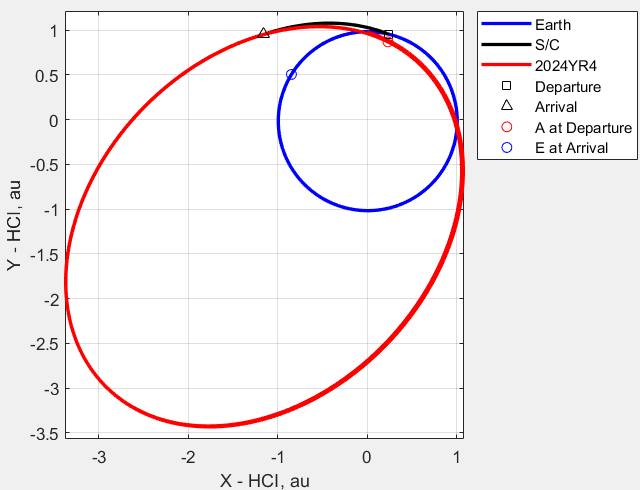}
  \caption{Exemplar 2024 YR$_{\text{4}}$ flyby reconnaissance trajectory using Janus with a December 2028 launch}
  \label{fig:janusDec2028}
\end{subfigure}%
\caption{Exemplar 2024 YR$_{\text{4}}$ flyby reconnaissance trajectories for the Janus spacecraft}
\label{fig:JanusTrajs}
\end{figure}

Indeed, this is what the trajectory analysis reveals. There is an approximately two-week launch period during the first half of June 2028 during which the spacecraft can launch and then encounter the asteroid approximately four months later, during mid-October 2028, with flyby speeds near 10 km/s and at a solar distance of 0.97 to 0.99 au. The launch $C_3$ varies from 42.188 to 51.801 km$^{\text{2}}$/s$^{\text{2}}$, so the launch vehicle would need to be a Falcon Heavy Recovery or Expendable, or a Vulcan Centaur 2 (VC2), VC4, or VC6. An exemplar trajectory is shown in Figure~\ref{fig:janusJun2028}, with launch $C_3$ of 45.539 km$^{\text{2}}$/s$^{\text{2}}$, Declination of Launch Asymptote (DLA) of -3.82$^{\circ}$, launch date of 2028-06-03, flight time of 136 days, arrival date of 2028-10-17, flyby speed of 9.92 km/s, and solar phase angle at flyby of 56.05$^{\circ}$.

A second, narrower launch opportunity occurs during the first week of December 2028. It requires very high energy launch ($C_3$ $>$ 94 km$^{\text{2}}$/s$^{\text{2}}$), necessitating a Falcon Heavy Expendable or VC6 launch vehicle despite the small mass of the Janus spacecraft. The flight times are short, between 70 and 86 days, the solar distance at asteroid encounter varies between 1.49 and 1.62 au, and the flyby speeds are relatively low, 1.62 to 1.85 km/s (viable for rendezvous using a larger spacecraft equipped with a sufficient propulsion system). An exemplar trajectory from this family of solutions is shown in Figure~\ref{fig:janusDec2028}, with launch $C_3$ of 99.135 km$^{\text{2}}$/s$^{\text{2}}$, DLA of 6.4$^{\circ}$, launch date of 2028-12-08, flight time of 72 days, arrival date of 2029-02-18, flyby speed of 1.79 km/s, and solar phase angle at flyby of 113.51$^{\circ}$. Although the low flyby speeds for this family of trajectories are favorable, the higher solar phase angles at flyby could reduce the quality of asteroid imaging, compared to flyby trajectories in the June 2028 launch period.

The time required to reassemble the Janus spacecraft, procure a suitable launch vehicle, and prepare the spacecraft for launch should be assessed to determine how far in advance of June 2028 the decision would need to be made to deploy the Janus spacecraft for reconnaissance of 2024 YR$_{\text{4}}$. The asteroid will not be visible to ground-based telescopes again until June 2028, but an attempt to use JWST to detect the asteroid in early 2026 is planned. If successful, that early 2026 detection should extend the observation arc sufficiently to significantly increase or decrease the lunar impact probability. That would enable the most informed decision about whether to deploy Janus to the asteroid. However, if JWST does not detect the asteroid in early 2026, then the decision about whether to deploy Janus in 2028 will have to be made with the lunar impact probability still at $\sim$4.3\%.

For any potential use of Janus to survey 2024 YR$_{\text{4}}$, additional work would be needed to analyze the suitability of the Janus payload and GNC system for the encounter. Janus was originally intended for flyby speeds of 3-3.5 km/s, of approximately $\sim$1 km diameter asteroids \cite{Shoer_2022}. Initial detection of 2024 YR$_{\text{4}}$ will be harder and the operational timeline will be substantially compressed for the much smaller object and likely faster flyby speed. It would also be necessary to understand whether the current payload would return sufficient information to be actionable, which may also be difficult given the small asteroid size and comparatively small imager carried by Janus~\cite{Chabot2024}.

\subsubsection{OSIRIS-APEX}

The NASA OSIRIS-APEX (Origins, Spectral Interpretation, Resource Identification and Security – Apophis Explorer) mission is a follow-on to the OSIRIS-REx mission, which launched in 2016 and returned samples of asteroid Bennu to Earth in 2023 \cite{Lauretta2017}. After the sample was returned, the spacecraft transitioned to the OSIRIS-APEX extended mission to rendezvous with the asteroid Apophis in June 2029, shortly after the asteroid’s close encounter with Earth in April 2029 \cite{Dellagiustina2023}. OSIRIS-APEX is scheduled to execute three Earth gravity assists before rendezvousing with Apophis, providing opportunities for redirecting the spacecraft for reconnaissance of 2024 YR$_{\text{4}}$ as shown in Table~\ref{tab_apex_events}~\cite{nolan2024osiris}. As of early 2024, the OSIRIS-APEX spacecraft had approximately 525 m/s of $\Delta$V remaining for maneuvers.

\begin{table}
    \centering
    \captionsetup{width=0.8\textwidth}
    \caption{OSIRIS-APEX Key Events}
    \begin{tabular}{lc}
        \toprule
        \textbf{Event} & \textbf{Date} \\
        \midrule
        Divert & 24 Sep 2023 \\
        DSM-1 (1 m/s) & 17 Jul 2024 \\
        EGA1 & 25 Sep 2025 \\
        DSM-2 (0.11 m/s) & 7 Oct 2026 \\
        EGA2 & 17 Mar 2027 \\
        DSM-3 (146 m/s) & 28 Jun 2027 \\
        EGA3 & 13 Apr 2029 \\
        Apophis Arrival & 22 Apr 2029 \\
        \bottomrule
    \end{tabular}
    \label{tab_apex_events}
\end{table}

The OSIRIS-APEX spacecraft was designed for small-body rendezvous and is equipped with thermal and infrared spectrometers, a laser altimeter, and multiple cameras. The camera suite is composed of three cameras: PolyCam for asteroid acquisition during approach and reconnaissance imaging, MapCam for recording color images during mapping, and SamCam for imaging the sample site and sample collection close to the asteroid. The OSIRIS-APEX guidance, navigation, and control (GNC) system as well as the camera suite is likely well suited for rendezvous reconnaissance of 2024 YR$_{\text{4}}$. However, a fast flyby of $\sim$60 m body such as 2024 YR$_{\text{4}}$ could present acquisition and terminal guidance challenges. Further evaluation of specific flyby scenarios is needed.

Multiple retasking scenarios of OSIRIS-APEX are considered, including rendezvous reconnaissance of 2024 YR$_{\text{4}}$, flyby reconnaissance of 2024 YR$_{\text{4}}$, and a dual flyby mission of both 2024 YR$_{\text{4}}$ and Apophis. Different diversion dates of the spacecraft from the OSIRIS-APEX nominal trajectory are evaluated for each scenario. Diverting from the nominal trajectory as late as possible is beneficial to allow more time for a retasking decision to be made. However, in general, the later the diversion date, the higher the $\Delta$V for achieving reconnaissance of 2024 YR$_{\text{4}}$. Critically, the nominal OSIRIS-APEX has a maneuver between each Earth gravity assist, targeting the flyby geometry to set up the subsequent gravity assist before the final EGA prior to the nominal rendezvous of Apophis. Adjusting the maneuver before the EGA allows for adjusting the flyby geometry for targeting 2024 YR$_{\text{4}}$ when the asteroid has its next close approach to Earth in late 2028.

\begin{figure}
	\centering
	\includegraphics[width=.8\textwidth]{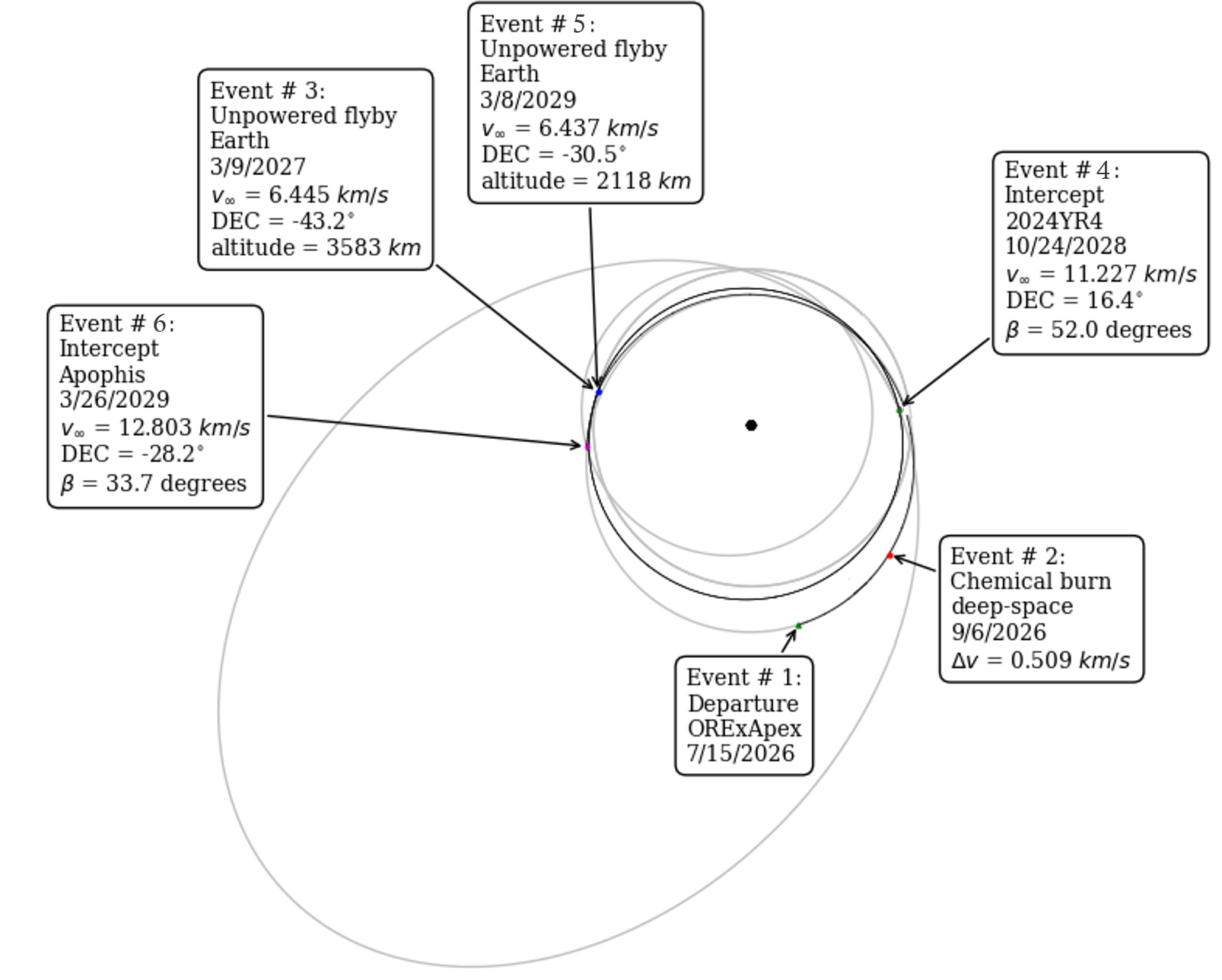}
	\captionof{figure}{Option D candidate trajectory to redirect OSIRIS-APEX to a fast flyby of 2024 YR$_{\text{4}}$ followed by a flyby of Apophis before its close approach to Earth}
	\label{fig_APEX_retask}
\end{figure}

Several possible flyby reconnaissance missions to 2024 YR$_{\text{4}}$ are identified and listed in Table~\ref{tab_yr4_apex_options}, adapted from~\cite{Bull_2025}. A 9.9 km/s flyby of 2024 YR$_{\text{4}}$ is feasible by retargeting EGA3 with only a one m/s maneuver on June 8, 2027 and skipping the nominal OSIRIS-APEX deep space maneuver later in June (Option A). The flyby speed can be reduced to as low as 8.2 km/s with a large deep space maneuver of roughly 510 m/s in early January 2027 (Option B). Reducing the arrival speed to a sufficiently low level for rendezvous with 2024 YR$_{\text{4}}$ with OSIRIS-APEX is not feasible. However, it is possible to fly by both 2024 YR$_{\text{4}}$ and Apophis with earlier diversion dates. Retargeting EGA2 with maneuver in November 2026 enables a trajectory with a 9.9 km/s flyby of 2024 YR$_{\text{4}}$ in early November 2028, and then a subsequent 11.2 km/s flyby of Apophis in September 2030 after its close approach with Earth (Option C). Such a mission would require 300 m/s of $\Delta$V. Alternatively, dual flyby mission with an 11.2 km/s flyby of 2024 YR$_{\text{4}}$ in October 2028 and a 12.8 km/s flyby of Apophis on March 26, 2029, before the Apophis close approach to Earth is viable for 510 m/s (Option D). The Option D trajectory is illustrated in Figure~\ref{fig_APEX_retask}, adapted from~\cite{Bull_2025}.

We performed a preliminary assessment of the ability of OSIRIS-APEX to detect 2024 YR$_{\text{4}}$ using the PolyCam instrument during terminal approach at 11.2 km/s relative speed (Option D). PolyCam specifications include an f-number of 3.5, focal length of 628.9 mm, pixel pitch of 8.5 microns, quantum efficiency of 0.31, post-tag transmission coefficient of 0.7626, and pointing stability of 0.0117 milliradians/sec (meeting science requirements referenced in~\cite{mission2015sawg}). The instrument has a wavelength sensitivity range of 370--1090 nm, and the detection criterion was set to SNR $\geq$7.0, with 600 millisecond exposures, each stacking 4 images. This results in a limiting visual magnitude of 14.87. In our modeling of the visual magnitude of the asteroid from the spacecraft's perspective during the approach trajectory, we use the H-G Slope formulation~\cite{dymock2007h}. Our modeling indicates that PolyCam would detect the 99.7\% HPDI Low Mass realization of 2024 YR$_{\text{4}}$ (the dimmest realization of the asteroid) 25.39 hours before intercept. For terminal phase course corrections, target detection at least 12 hours before intercept is required. Handling later detection times may be possible in certain circumstances, but autonomous operations might be required. Further analysis would be needed to determine whether OSIRIS-APEX could truly perform a successful flyby of 2024 YR$_{\text{4}}$, but this early assessment indicates potential feasibility.

\begin{table}[htbp]
    \centering
    \caption{Options for Retasking OSIRIS-APEX to 2024 YR$_{\text{4}}$}
    \setlength{\tabcolsep}{4pt} 
    \begin{tabular}{lcccccc}
        \toprule
        \textbf{Option} & \textbf{\begin{tabular}[c]{@{}c@{}}Diversion\\Date\end{tabular}} & \textbf{\begin{tabular}[c]{@{}c@{}}2024 YR$_{\text{4}}$\\Flyby Date\end{tabular}} & \textbf{\begin{tabular}[c]{@{}c@{}}2024 YR$_{\text{4}}$ Flyby\\ Speed (km/s)\end{tabular}}  & \textbf{\begin{tabular}[c]{@{}c@{}}Apophis\\Flyby Date\end{tabular}} & \textbf{\begin{tabular}[c]{@{}c@{}}Apophis Flyby\\ Speed (km/s)\end{tabular}} 
        & \textbf{\begin{tabular}[c]{@{}c@{}}Total\\$\Delta$V (m/s)\end{tabular}}\\
        \midrule
        \textbf{A} & 08 Jun 2027 & 01 Nov 2028 & 9.9  & N/A & N/A & 1 \\
        \textbf{B} & 02 Jan 2027 & 07 Nov 2028 & 8.2 & N/A & N/A & 510 \\
        \textbf{C} & 17 Nov 2026 & 01 Nov 2028 & 9.9  & 18 Sep 2030  & 11.2 & 300 \\
        \textbf{D} & 07 Sep 2026 & 24 Oct 2028 & 11.2  & 26 Mar 2029 & 12.8 & 510 \\
        \bottomrule
    \end{tabular}
    \label{tab_yr4_apex_options}
\end{table}

\subsubsection{Psyche}

The NASA Psyche mission launched in October 2023 for a planned rendezvous with the M-type asteroid Psyche in July 2029. Unlike OSIRIS-APEX it does not have any close approaches with Earth after launch, leaving Earth's vicinity to a semi-major axis of 2.92 au with a 21-kW solar electric propulsion (SEP) system and a Mars gravity assist (MGA). While Psyche does not have any EGAs to exploit for redirection towards 2024 YR$_{\text{4}}$, its SEP engines are propellant efficient and it launched with approximately 1100 kg of propellant, allowing for substantial maneuver capability. Psyche is equipped with a spectrometer, a magnetometer, and the Psyche Multispectral Imager (PMI). Given that the spacecraft was designed for rendezvous and not a flyby, further analysis of any flyby scenario is needed to fully understand retasking feasibility.

The long, continuous thrust arcs associated with Psyche's low-thrust SEP transfers create challenges for retasking evaluation. The remaining propellant load is dynamic during thrusting periods and must be considered in addition to the position and velocity state at the diversion date. To generate retasked trajectories, solutions are constrained to start along the nominal trajectory as defined in the predicted spacecraft ephemeris available on the NASA Navigation and Ancillary Information Facility (NAIF) website. However, the predicted propellant usage profile is not included with the ephemeris. Without the propellant available at any time in the nominal trajectory, the initial wet mass of the diverted spacecraft is assumed to be 2800 kg, which is the approximate wet mass of the spacecraft of launch. This assumption ensures the estimated propellant for the retasking segment of the trajectory is conservative. For the purposes of this study all four of Psyche's SPT-140 Hall thrusters are allowed to operate simultaneously with 90\% duty cycle. The nominal trajectory reserves one of the thrusters as a backup. Additionally, the beginning of life power from the solar arrays is assumed to be 21 kW with 1\% yearly degradation. The assumed power reserved for the bus during thrusting periods is 900 W.

\begin{figure}[h]
\centering
\begin{subfigure}{.47\textwidth}
  \centering
  \includegraphics[width=\linewidth]{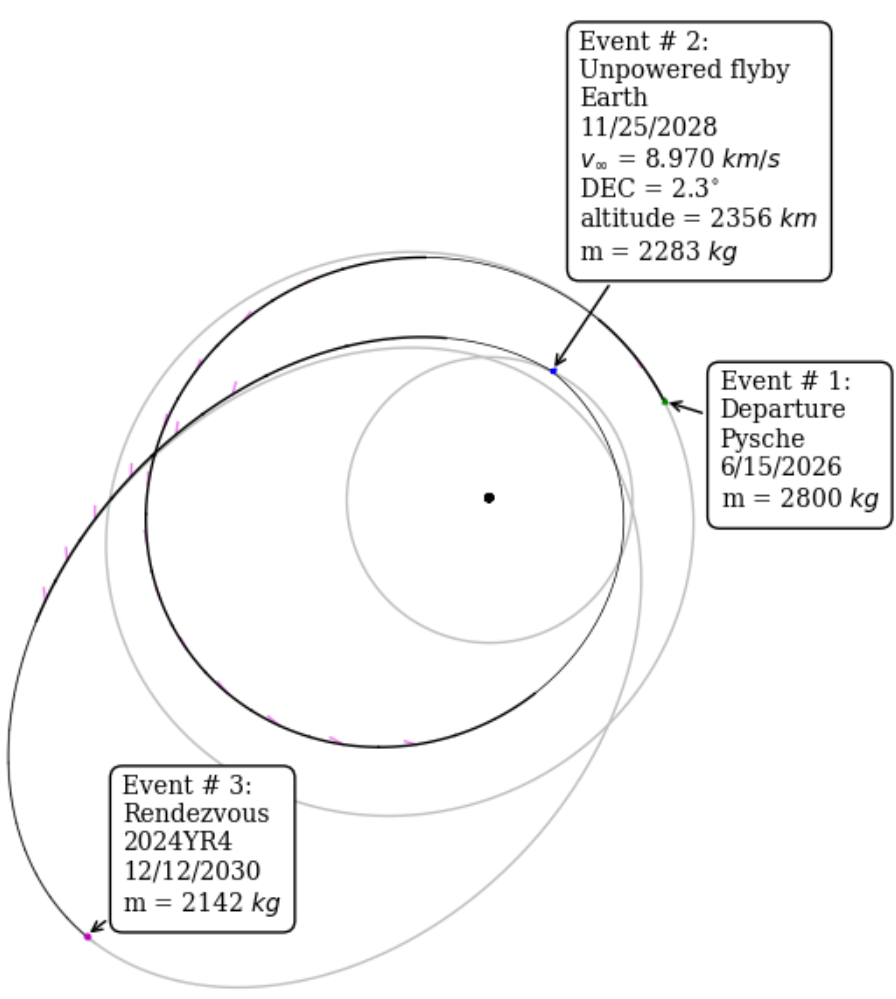}
  \caption{Candidate trajectory to redirect Psyche spacecraft to rendezvous with 2024 YR$_{\text{4}}$}
  \label{fig:psyche_rendezvous}
\end{subfigure}%
\hfill
\begin{subfigure}{.47\textwidth}
  \centering
  \includegraphics[width=\linewidth]{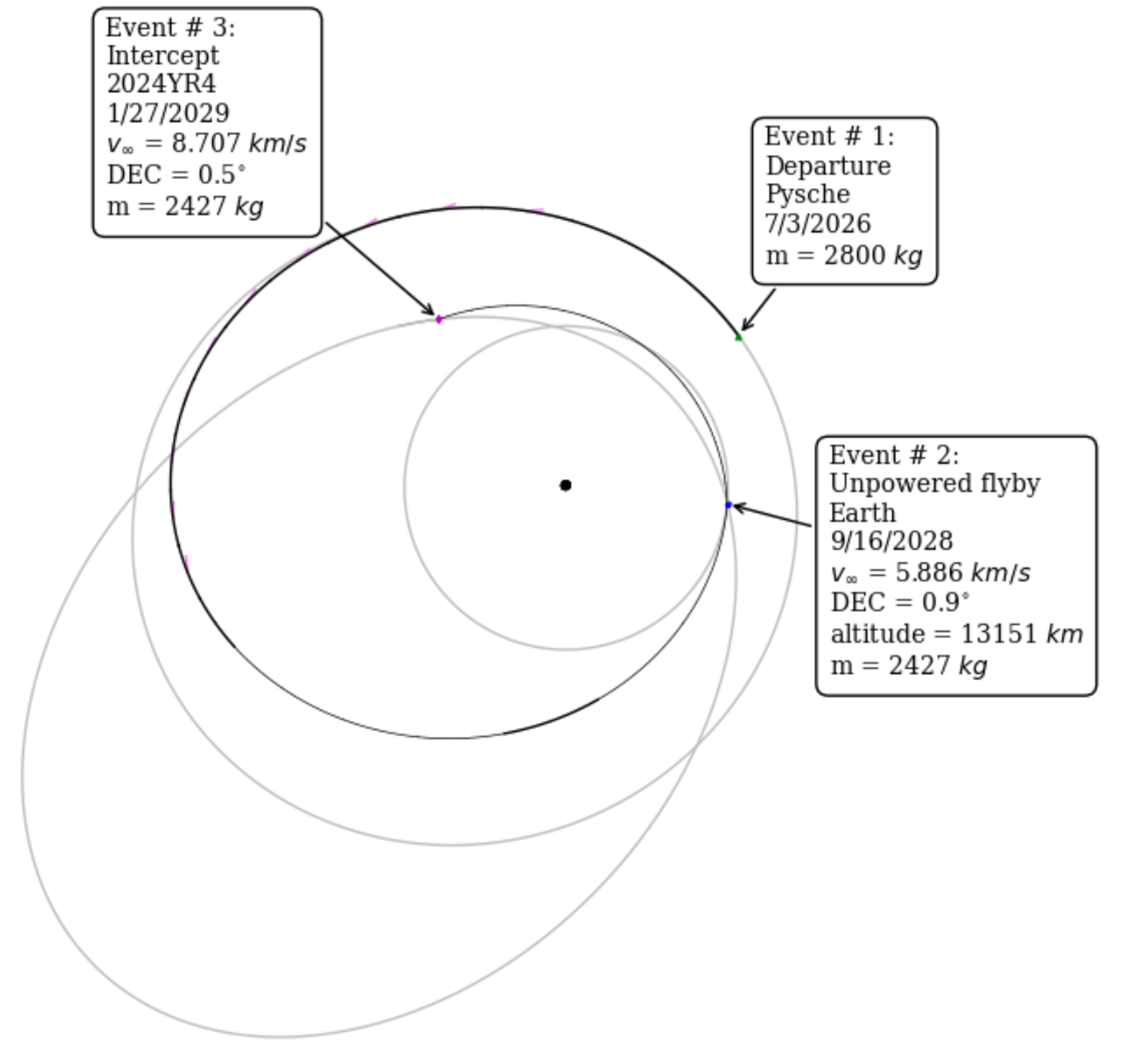}
  \caption{Candidate trajectory to redirect Psyche spacecraft to flyby with 2024 YR$_{\text{4}}$}
  \label{fig:psyche_flyby}
\end{subfigure}%
\caption{Exemplar 2024 YR$_{\text{4}}$  reconnaissance trajectories for the Psyche spacecraft}
\label{fig:PsycheTrajs}
\end{figure}

Two potential mission options for retasking Psyche for reconnaissance for 2024 YR$_{\text{4}}$ are identified, requiring diversion dates in mid-2026 after the MGA as highlighted in Table~\ref{tab:tab_pysche_retasking}. A rendezvous with 2024 YR$_{\text{4}}$ in December 2030 may be viable, but would require up to 725 kg of propellant starting in late June 2026 given the conservative assumption of a maximum starting wet mass of 2800 kg (Option A). The required propellant for this option may be more than is available at proposed diversion date given the nominal mission, necessitating further evaluation. The required propellant for rendezvous is roughly 640 kg if the initial mass of the spacecraft is 2500 kg at the diversion date. Alternatively, an 8.7 km/s flyby of 2024 YR$_{\text{4}}$ could be achieved in early 2029 with roughly 410 kg of propellant (Option B). Trades between arrival date, flyby speed, and required propellant are possible for different 2024 YR$_{\text{4}}$ flyby scenarios. Both Option A and B trajectories are plotted in Figure \ref{fig:PsycheTrajs}.

\begin{table}[htbp]
    \centering
    \caption{Psyche Retasking Options}
    \setlength{\tabcolsep}{3pt} 
    \begin{tabular}{l>{\centering\arraybackslash}p{1.8cm}>{\centering\arraybackslash}p{1.8cm}>{\centering\arraybackslash}p{1.8cm}>{\centering\arraybackslash}p{1.8cm}p{3.2cm}}
        \toprule
        \textbf{Option} & \textbf{Diversion Date} & \textbf{Arrival Date} & \textbf{Flyby Speed [km/s]} & \textbf{Required Propellant [kg]*} & \textbf{Notes} \\ 
        \midrule
        \textbf{A} & 15 Jun 2026 & 12 Dec 2030 & Rendezvous & $\sim$725 kg & EGA on 25 Nov 2028, Rerouting occurs after baseline MGA \\ 
        \textbf{B} & 03 Jul 2026 & 27 Jan 2029 & 8.7 km/s & $\sim$410 kg & EGA on 16 Sep 2028, Rerouting occurs after baseline MGA \\ 
        \bottomrule
        \multicolumn{6}{l}{\small *Estimated required propellant includes 10\% margin and assumes initial mass of 2800 kg} \\
    \end{tabular}
    \label{tab:tab_pysche_retasking}
\end{table}

\subsection{Purpose-Built Flyby and Rendezvous Reconnaissance Missions}

\begin{figure}[h]
    \centering
    \includegraphics[width=0.7\textwidth]{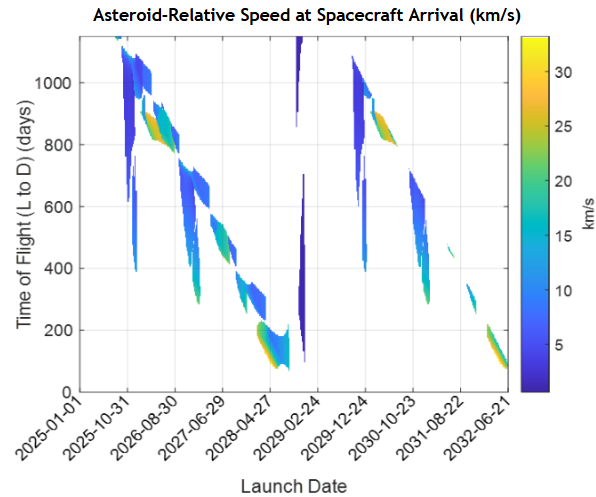}
    \caption{Pork Chop Contour (PCC) plot of relative velocity at asteroid arrival for ballistic trajectories across the ranges of launch dates and times of flight considered for 2024 YR$_{\text{4}}$.}
    \label{fig:PCCBallAll}
\end{figure}

\subsubsection{Rapid Response Reconnaissance}

We ran a ballistic trajectory scan with launch dates as early as 2025-01-01, to search for rapid response reconnaissance launch opportunities that could have been utilized shortly after 2024 YR$_{\text{4}}$ was discovered in late December 2024 and recognized as a potential Earth impactor. No such rapid response reconnaissance capability currently exists, so this analysis was for purposes of illustrating the utility of such capability.

\begin{figure}[h]
\centering
\begin{subfigure}{.45\textwidth}
  \centering
  \includegraphics[width=\linewidth]{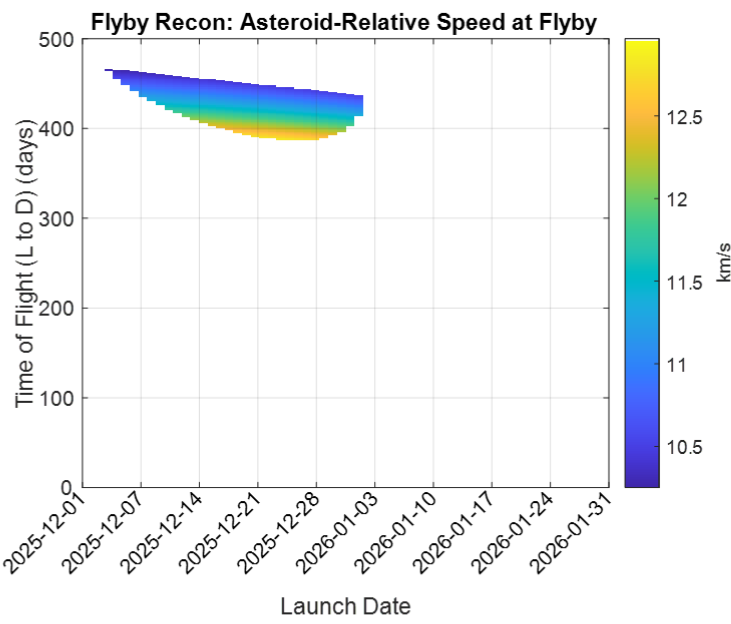}
  \caption{PCC plot showing rapid response launch period in December 2025 for 2024 YR$_{\text{4}}$.}
  \label{fig:rapidPCC}
\end{subfigure}%
\hfill
\begin{subfigure}{.45\textwidth}
  \centering
  \includegraphics[width=\linewidth]{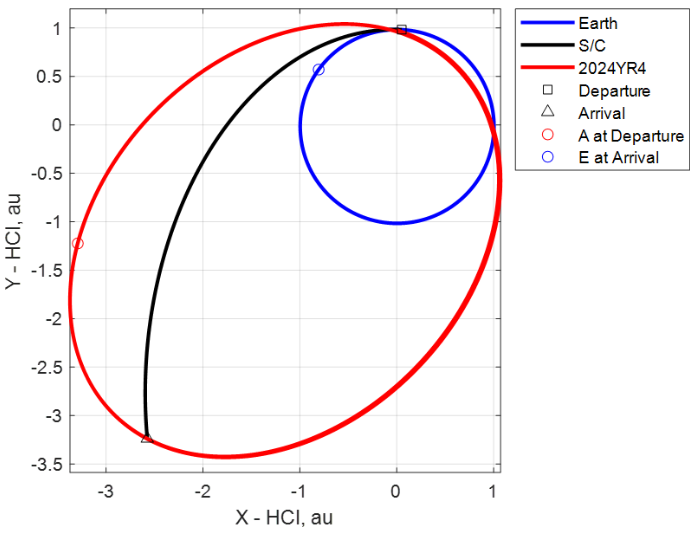}
  \caption{Exemplar 2024 YR$_{\text{4}}$ rapid response flyby reconnaissance trajectory launching December 2025.}
  \label{fig:rapidTraj}
\end{subfigure}%
\caption{PCC plot of rapid response flyby reconnaissance trajectories in December 2025 for 2024 YR$_{\text{4}}$ and exemplar trajectory from that set.}
\label{fig:JanusTrajs}
\end{figure}

Examination of Figure~\ref{fig:PCCBallAll} reveals that the trajectories with the earliest asteroid arrival dates launch in late 2025. Figure~\ref{fig:rapidPCC} presents a PCC plot focused on that family of trajectories, which have launch dates spanning most of December 2025 and reach the asteroid between January and March of 2027. This would provide confirmation of whether the asteroid is actually on a lunar impact trajectory $\sim$16 months sooner than waiting for the asteroid to become visible to ground-based telescopes again in June 2028. It would also provide flyby physical characterization data for the asteroid much sooner than either the Janus option or a non-rapid response purpose-built option. That earlier knowledge of the asteroid's physical properties would remain valuable even if the currently planned early 2026 JWST observation of the asteroid is able to provide even earlier confirmation of whether the asteroid is on a lunar impact trajectory.

However, it should be noted that even if a rapid response reconnaissance capability had been available, the nature of the asteroid's orbit is such that the asteroid would not have been reachable right away. The earliest viable launch opportunity is about 11 months after the asteroid was discovered, and the reconnaissance spacecraft could not reach the asteroid until just over two years after it was discovered. A rapid response capability is generally assumed to be able to launch within, at most, a few months after the asteroid is discovered and then gather data on the asteroid within a few months after launch. 2024 YR$_{\text{4}}$ provides a real-life example of an asteroid orbit for which more capable rapid response systems would be needed to reach it this quickly, in particular higher launch $C_3$ capabilities than the maximum of 100 km$^{\text{2}}$/s$^{\text{2}}$ currently offered by the Falcon Heavy Expendable and planned to be offered by the VC6. Alternatively, discovering Earth- (or Moon-) impacting asteroids farther in advance of their impact dates is another way to enable timely reconnaissance. Indeed, NASA's forthcoming Near-Earth Object (NEO) Surveyor space-based telescope~\cite{Mainzer_2023} is designed to find potentially hazardous objects farther in advance, to provide us with increased warning time.

\subsubsection{Purpose-Built Flyby and Rendezvous Reconnaissance}

If allowing for a more traditional spacecraft development timeline, at least three years would be allotted from authority to proceed (ATP) until launch. We examined launch dates no earlier than January 2028 and arrival dates up to the 2024 YR4 close approach with Earth in December 2032. Later launches provide more time for spacecraft development and earlier arrival times are preferred so that critical asteroid information for any mitigation is acquired with time to adjust the mitigation plan and execution. Both flyby and rendezvous reconnaissance are evaluated. While a flyby mission can potentially allow for a faster development period and shorter flight times than rendezvous, rendezvous enables improved characterization. 

\begin{figure}
    \centering
    \includegraphics[width=.95\textwidth]{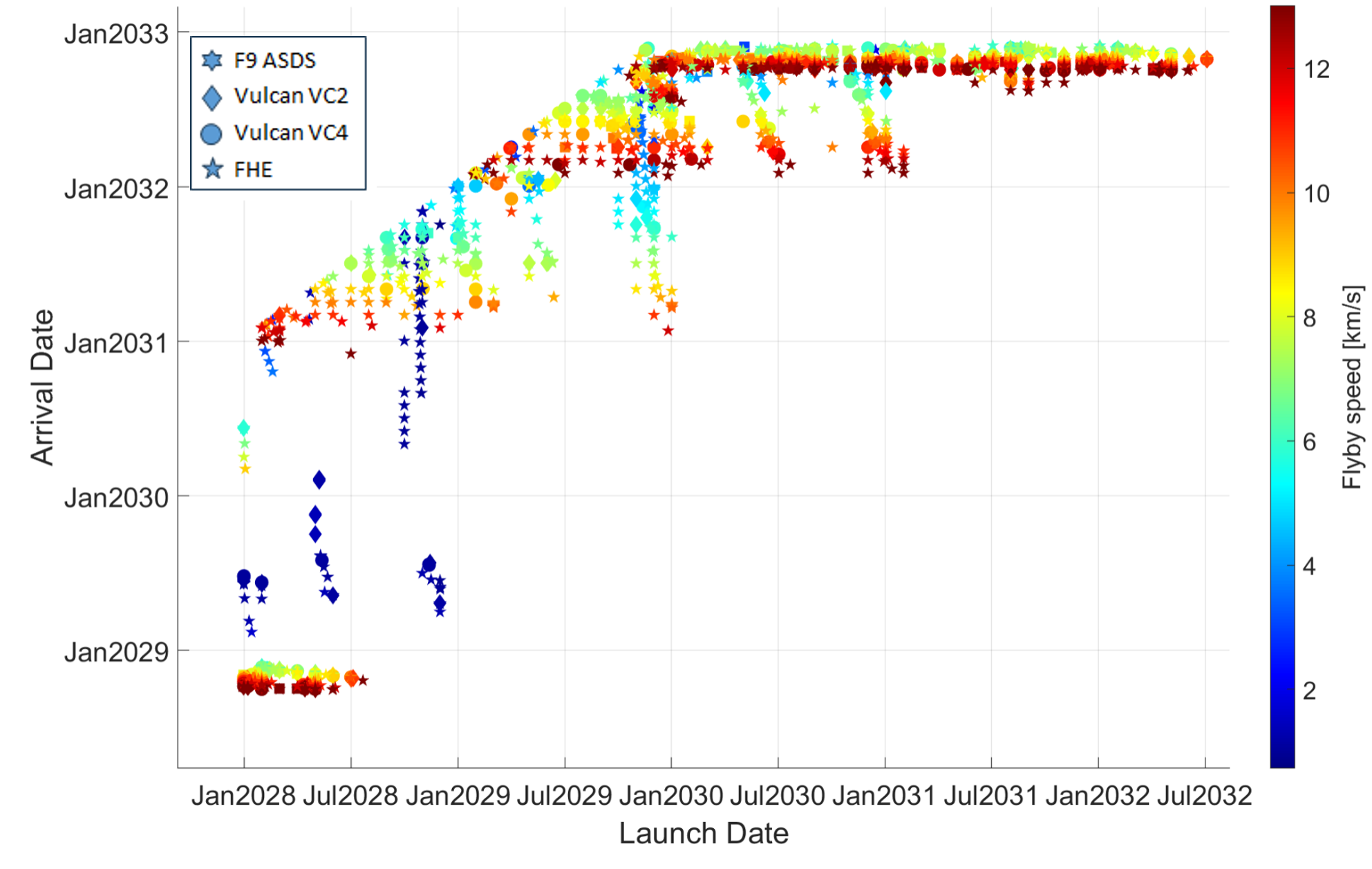}
    \captionof{figure}{Pareto optimal solution space for flyby reconnaissance missions of 2024 YR$_{\text{4}}$ when minimizing launch vehicle class, maximizing delivered mass, minimizing flyby speed, maximizing launch date, and minimizing arrival date. All solutions have a delivered mass greater than 500 kg.}
    \label{fig:ChemFlybyParetoFront}
\end{figure}

A broad set of mission design parameters are traded for flyby reconnaissance missions with chemical propulsion. We evaluated different launch vehicle classes spanning from small- to heavy-lift vehicles for interplanetary missions: Falcon 9 (F9) Autonomous Spaceport Drone Ship (ASDS), a Vulcan VC2, a Vulcan VC4, and a Falcon Heavy Expendable. Up to two gravity assists from Venus, Earth or Mars are allowed to reduce propellant demand. Additionally, a deep space maneuver of up to 2 km/s between any two bodies is allowed with an assumed specific impulse, Isp, of 320 s. Figure \ref{fig:ChemFlybyParetoFront} depicts Pareto optimal solutions with at least 500 kg of delivered mass given equally weighted objectives of: minimizing launch vehicle class, maximizing delivered spacecraft mass, minimizing flyby speed, maximizing launch date, and minimizing arrival date. Some families of solutions require high $\Delta $V from the spacecraft propulsion system, driving high propellant mass fractions. Traditional spacecraft typically have propellant mass fractions below 60\%. Missions with a higher propellant mass fraction would likely require staging or be infeasible from a spacecraft hardware perspective.

\begin{figure}
    \centering
    \includegraphics[width=.7\textwidth]{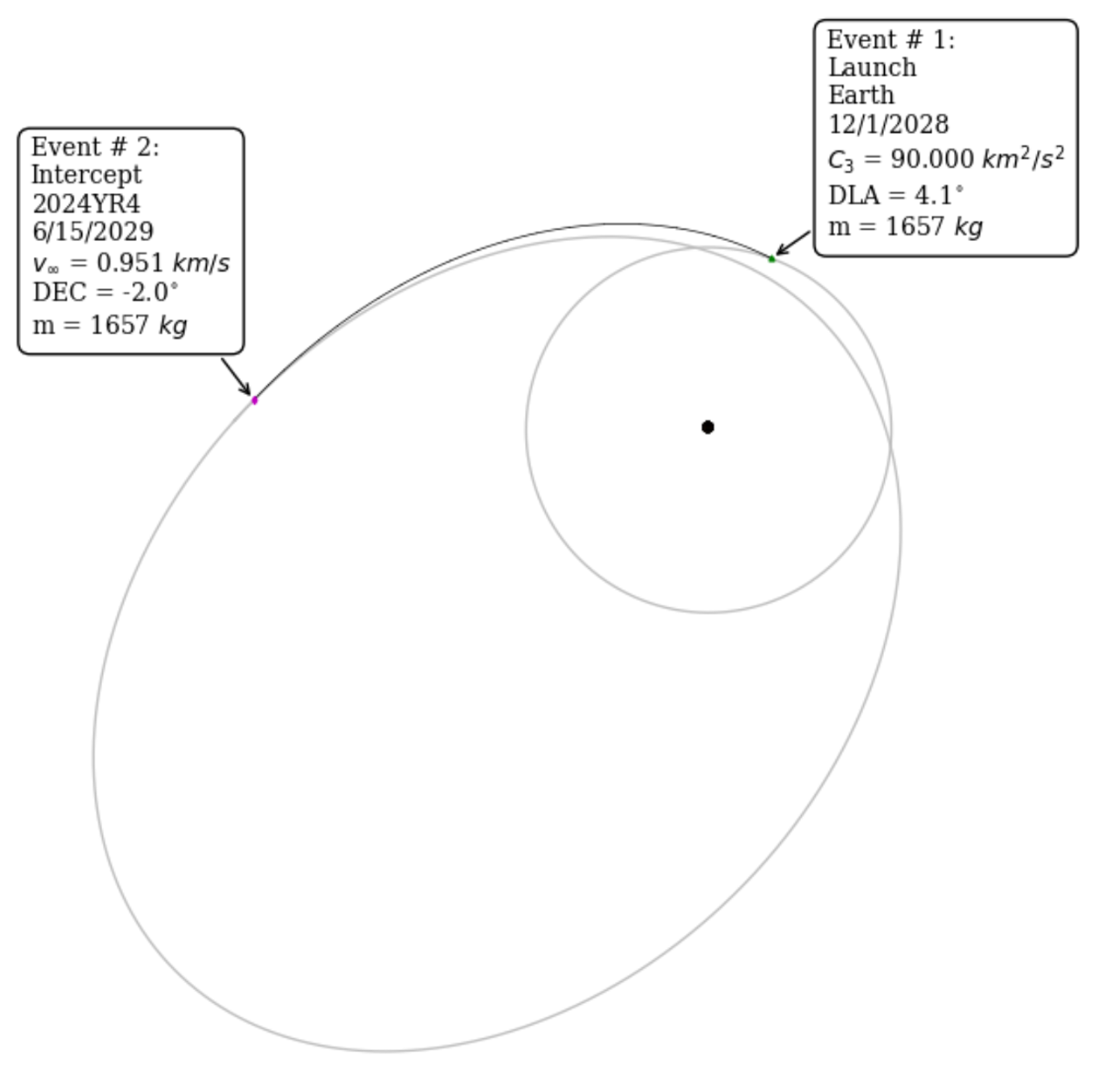}
    \captionof{figure}{Exemplar flyby reconnaissance trajectory to 2024 YR$_{\text{4}}$, launch in December 2028 and flying by the asteroid in June 2029.}
    \label{fig:ChemFlybyExample}
\end{figure}

The trade study results reveal several distinct solution families based on launch and flyby timing. Notably, the solution space shows that a launch in 2028 is required to enable a low speed flyby (less than 4 km/s) with an arrival date before 2031. As an alternative, flybys in 2031 are feasible, but the spacecraft must launch by late 2029, and the lowest possible flyby speed is $\sim$5 km/s. If a five-year development duration is required with a launch no earlier than mid- to late-2030, the earliest viable flyby date is early 2032 with flyby speeds greater than 10 km/s. An example trajectory from the family of low-speed flyby solutions with a launch in December 2028 and arrival in mid-2030 is plotted in Figure~\ref{fig:ChemFlybyExample}.

We performed a preliminary assessment of the ability of our notional flyby reconnaissance mission spacecraft to detect 2024 YR$_{\text{4}}$ during approach using the DRACO instrument employed by NASA's DART spacecraft to successfully detect and impact the asteroid Dimorphos. DRACO has an f-number of 12.6, a focal length of 2620 mm, and pixel pitch of 6.5 microns, featuring an estimated integrated quantum efficiency of 0.3388 and optics transmission coefficient of 0.7614. It demonstrates read noise of 2 e-/pix/exposure, dark current of 35 e-/pix/s~\cite{DRACO_SIS_2022}, SNR limit of 173, and exceptional pointing stability of 0.002 milliradians/sec (1$\sigma$). DRACO has a wavelength sensitivity range of 370--1090 nm, and the detection criterion was set to SNR $\geq$7.0, with 300 millisecond exposures, each stacking 4 images. The resulting limiting visual magnitude was calculated to be 16.13. As with the prevously reported flyby analysis for OSIRIS-APEX, in our modeling of the visual magnitude of the asteroid from the spacecraft's perspective during the approach trajectory, we use the H-G Slope formulation~\cite{dymock2007h}. Our modeling for the low-speed reconnaissance flyby trajectory in~\ref{fig:ChemFlybyExample} indicates that the DRACO instrument would detect the 99.7\% HPDI Low Mass realization of 2024 YR$_{\text{4}}$ (the dimmest realization of the asteroid) approximately 137 hours before closest approach. For terminal phase course corrections, target detection at least 12 hours before intercept is required. Handling later detection times may be possible in certain circumstances, but autonomous operations might be required. Further analysis would be needed to determine whether a flyby reconnaissance mission employing DRACO or a DRACO-like terminal guidance sensor could truly perform a successful flyby of 2024 YR$_{\text{4}}$, but this early assessment indicates potential feasibility.

\begin{figure}
    \centering
    \includegraphics[width=.95\textwidth]{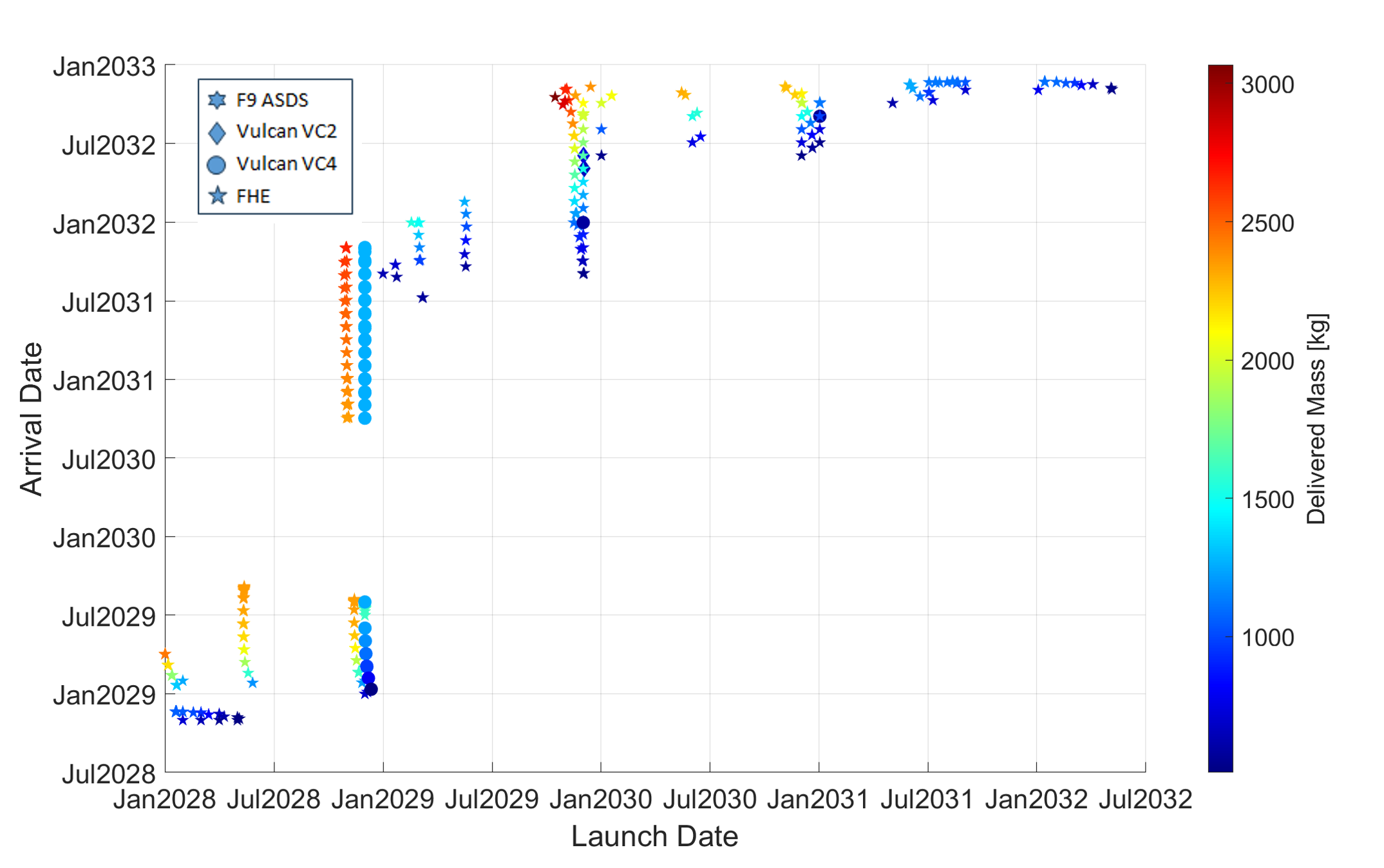}
    \captionof{figure}{Pareto optimal solution space for chemical propulsion based rendezvous when maximizing delivered mass, maximizing launch date, and minimizing arrival date. All solutions shown have a delivered mass greater than 500 kg}
    \label{fig:ChemRendezvousParetoFront}
\end{figure}

Fewer chemical-based rendezvous reconnaissance options exist than flyby options given the high $\Delta$V demand of most arrival maneuvers. The Pareto optimal solution space for a purpose-built rendezvous spacecraft with chemical propulsion is illustrated in Figure \ref{fig:ChemRendezvousParetoFront}. The same mission design trade parameters are considered for rendezvous as for the flyby reconnaissance with up to two gravity assists and deep space maneuvers less than 2 km/s between bodies. Mission designs are optimized for equally weighted objective functions of maximizing delivered mass, maximizing launch date, and minimizing arrival date. The latest possible launch with arrival before 2032 is December 2029. Ideally, a reconnaissance rendezvous mission would launch in November or December 2028 to allow for 2029 arrival dates and the ability to modify any mitigation mission before launch. As with the chemical flyby trajectories, some solutions are associated with high $\Delta$V and high propellant mass fractions. Solutions with a delivered mass below 1000 kg in Figure \ref{fig:ChemRendezvousParetoFront} are often associated with high propellant mass fractions and would require further analysis to understand development feasibility. All solutions with a launch later than January 2029 and arrival earlier than January 2032 require propellant mass fractions greater than 75$\%$. Trajectories with a launch later than January 2029 and lower propellant mass fractions become viable with arrival dates in 2032. As an example, a trajectory launching on a Falcon Heavy Expendable in November 2029 and an arrival in May 2032 would have a 65$\%$ propellant mass fractions and be capable of delivering over 1800 kg after the rendezvous maneuver.

\begin{figure}
    \centering
    \includegraphics[width=.95\textwidth]{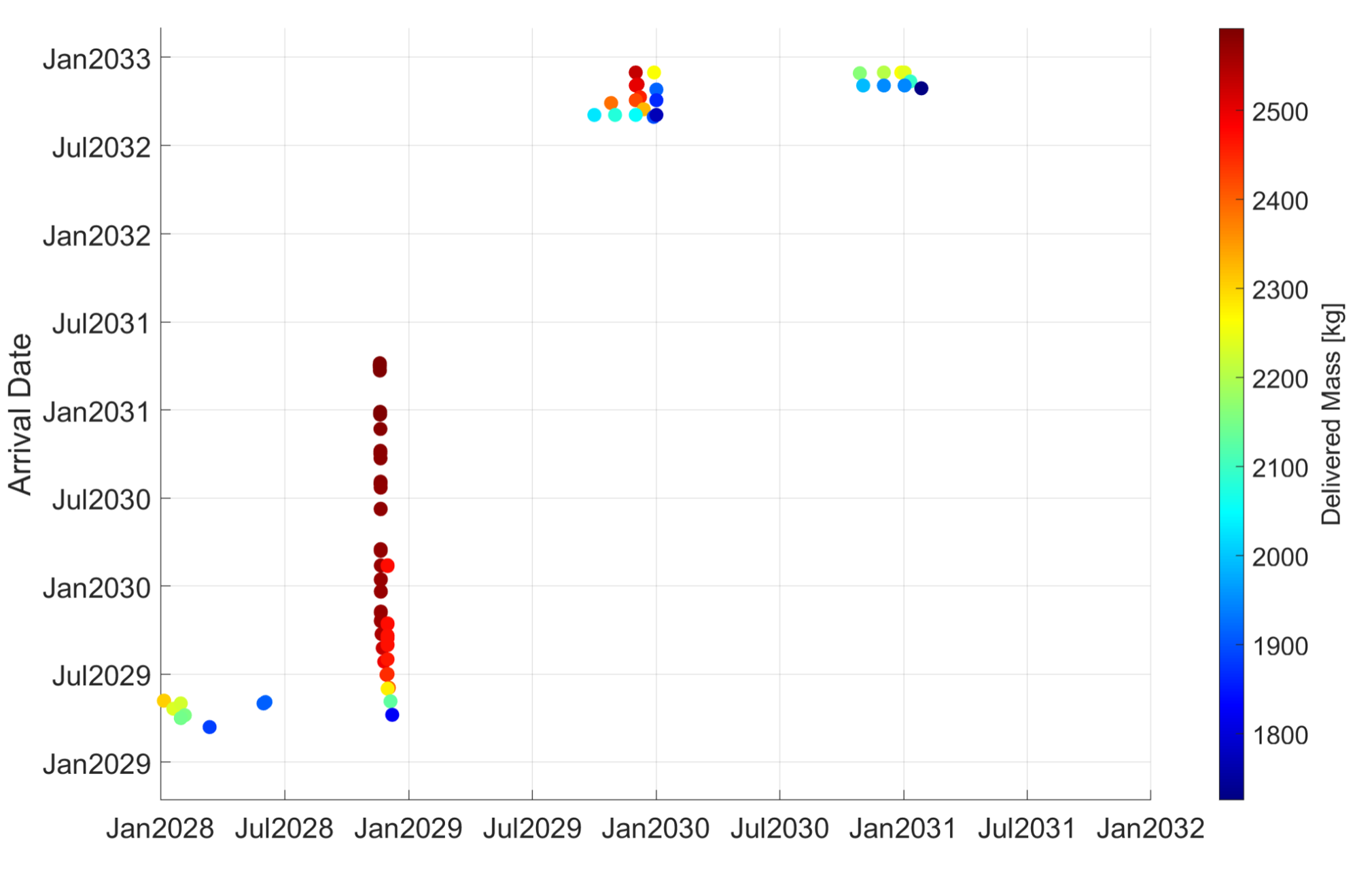}
    \captionof{figure}{Pareto optimal solution space for trajectories with delivered mass greater than 1700 kg when maximizing delivered mass, maximizing launch date, and minimizing arrival date for SEP rendezvous missions to 2024 YR$_{\text{4}}$.}
    \label{fig:SEPRendezvousParetoFront}
\end{figure}

A SEP-based spacecraft can be beneficial for high $\Delta$V scenarios, but the need for a fast development timeline must also be weighed. A rebuild of the Psyche spacecraft design is evaluated with the baseline 21-kW solar electric propulsion system and four SPT-140 thrusters. Rebuilding to the same Psyche design would ideally save development time compared to a fully-customized spacecraft design. The Pareto front of optimal solutions for a Psyche-based SEP spacecraft to rendezvous with 2024 YR4 is plotted in Figure \ref{fig:SEPRendezvousParetoFront}. The solution space for the SEP spacecraft is somewhat sparse when filtered for trajectories that can deliver at least 1700 kg, the approximate Psyche dry mass. The same late-2028 launch opportunity as observed in the chemical flyby and rendezvous solution space is prominent with arrival dates from early 2029 to mid 2031. Additionally, there are several earlier launch opportunities in 2028 (January to June) with mid-2029 rendezvouses with 2024 YR4. These early 2028 launches with a SEP system may be more attractive than chemical trajectories with similar launch and arrival dates as the SEP system provides much more favorable propellant mass fractions. 

\subsection{Kinetic Impactor Deflection Mission Analysis}

A Lambert grid scan was performed to evaluate candidate KI trajectories. Partial derivatives of the asteroid's coordinates on the Earth's B-plane with respect to the components of the applied $\Delta V$ vector were utilized to calculate the deflection achieved by each KI trajectory. KI mass was limited by the standard launch mass vs. $C_3$ curve for the SpaceX Falcon Heavy Expendable launch vehicle. A code was developed that optimally selects sequences of KI mission trajectories and adjusts each KI's mass such that each KI strike on the asteroid produces a $\Delta V$ that will be sufficient for deflection if $\beta = 2$ while not exceeding the 10\% $V_{\text{escape}}$ threshold for unwanted fragmentation if $\beta = 5$. When even the most effective such KI strike on the asteroid does not produce sufficient deflection, the code will then apply multiple KI strikes, in descending order of the most effective trajectory solutions, until sufficient deflection is achieved. In this way, the minimum number of KI spacecraft are used to achieve full deflection while keeping each $\Delta V$ on the asteroid no larger than 10\% $V_{\text{escape}}$ if $\beta$ turns out to be as high as 5. While $\beta$ could theoretically exceed 5 in some circumstances, 5 is currently regarded as a reasonable generic upper limit, based on the DART data~\cite{Cheng_2023}.

This code was applied to 2024 YR$_{\text{4}}$ to calculate the minimum set of KI deflection missions for the 99.7\% Low Mass, 50th percentile mass, and 99.7\% High Mass realizations of the asteroid, and for each of those cases full chord, half chord, and 10\% chord deflection distances were considered. For the first analysis, the earliest allowable launch date was set to 2028-06-01. For the 99.7\% High Mass asteroid, a single 72 kg KI launching on 2028-06-09 and striking the asteroid at 9.9 km/s on 2028-11-14 is sufficient to deflect the asteroid for deflection distances up to $\sim$80\% of the impact risk chord length, while staying below the fragmentation threshold. Deflecting the asteroid for the full chord length would require a second KI of at least 17 kg mass, launched on 2028-06-19 and striking the asteroid at 9.65 km/s on 2028-10-31.

The situation is similar for the 50th percentile mass realization of the asteroid, but the KI masses are smaller, ~16--17 kg. The situation for the 99.7\% Low Mass realization of the asteroid is quite different, however. It could be deflected up to $\sim$30\% of the chord length by a very small $\sim$2 kg KI using either of the two trajectories described above. Deflecting it 50\% of the chord length would require two such KIs. Deflecting it the full chord length would require 16 such small KIs with launch dates spanning June 2028 up to April 2032, completing the deflection 3 months before the lunar encounter date. If launching as early as 2026-04-08 were possible, the number of KIs could be reduced from 16 to 6. This is mentioned for completeness---neither of those approaches with so many small KIs is considered practical. In fact, the question of whether splitting a deflection up into multiple impulses to keep each under the fragmentation threshold will actually avoid unwanted fragmentation remains unanswered. Further modeling and simulation work is needed to properly address that question.

Being able to launch by July 2028 is critical for being able to efficiently deflect the asteroid via KIs, because it enables trajectories that strike the asteroid at or near its perihelion in November 2028. Deflection performance falls off steeply if the earliest launch is not possible until sometime between August and October of 2028. If launch is not possible until after October 2028, it would require an impractical number of KIs (dozens) to deflect the asteroid. This is because deflection performance decreases rapidly as the asteroid travels farther and farther away from perihelion, and the available intercept geometries deviate from asteroid velocity vector aligned intercepts.

These results illustrate the fact that the current uncertainties in the asteroid's mass and lunar impact location (should it turn out to indeed be on a lunar impact trajectory) make it impossible to specify a KI spacecraft mission design that we would be confident would have sufficient capability to deflect the asteroid without causing unwanted fragmentation and potentially generating debris that could adversely affect crew or space assets in cislunar space or Earth orbit. A reconnaissance mission would therefore be necessary to reduce the uncertainties in the asteroid's mass and impact location sufficiently to inform the design of a KI deflection mission. 

The need for a reconnaissance mission prior to the design and construction of a KI spacecraft poses a significant challenge in this situation because it would already be quite challenging to have KI spacecraft ready to launch by June or July 2028 even if design and construction were started now (August 2025). That leaves no time to design, build, and fly a reconnaissance mission that could inform the design of a KI spacecraft needing to launch in June or July 2028.

For these reasons, we conclude that KI deflection of 2024 YR$_{\text{4}}$ is not practical.

\subsection{Nuclear Deflection Mission Analysis}

Opportunities for NED deflection of 2024 YR$_{\text{4}}$ are reckoned by observing the points in time when the required deflection $\Delta V$ crosses the unwanted fragmentation thresholds (10\% $V_{\text{escape}}$) in Figure~\ref{fig:Defl_DV_zoom}. The largest $\Delta V$ that can be applied to each reference realization of the asteroid, equal to 10\% $V_{\text{escape}}$, is given in Table~\ref{tab:NED_defl_DV}, along with the approximate HOBs needed to achieve those $\Delta V$s using a 100 kt NED, which are calculated using the aforementioned analytical model~\cite{Managan_2025}.

\begin{table}[htbp]
\centering
\caption{Maximum single-impulse deflection $\Delta V$ Magnitudes for 2024 YR$_{\text{4}}$ and HOBs w/ 100 kt NED}
\label{tab:NED_defl_DV}
\begin{tabular}{lccc}
\toprule
 & 99.7\% Low Mass & 50th \%tile Mass & 99.7\% High Mass \\
\midrule
$\Delta V$        & 0.18 cm/s & 0.336 cm/s & 0.504 cm/s  \\
HOB w/ 100 kt NED & 560 m     & 550 m      & 530 m       \\
\bottomrule
\end{tabular}
\end{table}

The dates when the required $\Delta V$ for deflection reaches the maximum allowable $\Delta V$ values in Table~\ref{tab:NED_defl_DV} are shown in Table~\ref{tab:NED_defl_dates}. These are the dates by which an NED must be detonated at the appropriate HOB above the asteroid's surface, aligned with the asteroid's heliocentric inertial velocity vector, to achieve the desired amount of deflection.

\begin{table}[htbp]
\centering
\caption{Latest Dates For Single-Impulse Deflection of 2024 YR$_{\text{4}}$}
\label{tab:NED_defl_dates}
\begin{tabular}{lccc}
\toprule
Deflection Distance & 99.7\% Low Mass & 50th \%tile Mass & 99.7\% High Mass \\
\midrule
Full Chord & N/A        & 2029-02-05 & 2029-06-17 \\
Half Chord & 2029-03-01 & 2029-10-19 & 2030-05-16 \\
10\% Chord & 2030-12-20 & 2031-06-13 & 2031-09-30 \\
\bottomrule
\end{tabular}
\end{table}

For purposes of our current analysis, we will assume that only one NED impulse will be applied to the asteroid for deflection. Similar to the discussion of KI deflection, we will assume that applying multiple smaller impulses to 2024 YR$_{\text{4}}$ to deflect it while remaining below the unwanted fragmentation threshold would not be worth the complexities that would entail, given that the asteroid is small enough to be readily robustly disrupted. Additionally, we will only consider rendezvous trajectory options for delivering NEDs to the asteroid for deflection. While it is possible that NEDs could be detonated precisely enough during high-speed intercept to achieve a targeted deflection $\Delta V$, the challenges associated with doing that would not be worth taking on when a kinetic or nuclear robust disruption of the asteroid could be done more simply instead.

We also assume that at least 2000 kg of total spacecraft mass must be delivered to rendezvous with the asteroid in order to contain two 100 kt free-flyer NEDs to the asteroid. This is a rough order of magnitude mass estimate that can be refined in future work beyond the scope of this paper. Two NEDs are included to provide redundancy. The concept of operations for NED deflection of the asteroid involves arriving 30 days before the desired deflection date, surveying the asteroid, selecting the HOB, deploying the free-flyer NED to station-keep at the correct detonation coordinates relative to the asteroid, moving the main spacecraft to the opposite side of the asteroid, detonating the NED, and then using the main spacecraft to continue monitoring the situation. A second NED is onboard in case it is needed, otherwise it can be safely disposed of by detonating it in deep space after the asteroid is successfully deflected by the first NED.

\begin{table}[htbp]
\centering
\caption{Chemical Propulsion Rendezvous Missions to 2024 YR$_{\text{4}}$}
\label{tab:NED_rndz_chem}
\begin{tabular}{llcccl}
\toprule
Launch  & Arrival & Delivered & Propellant & Prop. Mass & Deflection  \\
Date & Date & Mass (kg) & Mass (kg) & Fraction & Case\\
\midrule
11/14/2028 & 5/16/2029 & 2246 & 3307 & 0.6 & High Mass, Full Chord \\
11/13/2028 & 6/15/2029 & 2311 & 3286 & 0.59 & High Mass, Full Chord \\
11/13/2028 & 8/6/2029 & 2344 & 3260 & 0.58 & $>$50th \%tile Mass, Half Chord \\
11/3/2028 & 11/5/2030 & 2367 & 2754 & 0.54 & Low Mass, 10\% Chord\\
10/29/2028 & 6/2/2031 & 2499 & 2805 & 0.53 & 50th \%tile Mass, 10\% Chord\\
10/28/2028 & 9/30/2031 & 2609 & 3063 & 0.54 & High Mass, 10\% Chord\\
\bottomrule
\end{tabular}
\end{table}

Table~\ref{tab:NED_rndz_chem} lists rendezvous trajectory options using chemical propulsion ($\sim$320 s I$_{\text{sp}}$) that can deliver at least 2000 kg of total spacecraft mass to the asteroid. A SEP option is also available using a Psyche-like spacecraft that launches on 2028-12-03 and rendezvouses with the asteroid on 2029-06-04, delivering a total of 2391 kg. These rendezvous options are able to perform deflection for all but three of the cases outlined in Table~\ref{tab:NED_defl_dates}: Low Mass with Full Chord or Half Chord, or 50th Percentile Mass with Full Chord.

While 6 of the 9 combinations of asteroid mass and deflection distance in Table~\ref{tab:NED_defl_dates} could be deflected by a rendezvoused NED, in principle, the launch dates are all in late 2028 and there are no later launch options for rendezvous that can deliver sufficient mass. That leaves only three years from now (August 2025, at the time of this writing) to build the spacecraft, but no decisions have yet been made. Considering that lunar impact is currently uncertain and that building an interplanetary spacecraft capable of rendezvous typically takes at least 5 years, we do not find NED deflection of 2024 YR$_{\text{4}}$ to be a practical option.

\subsection{Kinetic Robust Disruption Missions}

The bright yellow regions in Fig.~\ref{fig:KI_disr_PCC} indicate the most performant mission trajectory opportunities for kinetic robust disruption. Those launch opportunities are available in late 2029 to mid 2030, late 2030, mid 2031, late 2031, and early to mid 2032. Most of those trajectory options involve disrupting the asteroid in August or September of 2032, a few months prior to the lunar encounter date. If 2024 YR$_{\text{4}}$ turns out to be around the median size/mass (or smaller), then Fig.~\ref{fig:KI_disr_PCC} indicates that there are additional launch opportunities that would disrupt the asteroid at significantly earlier dates and provide much more time for post-disruption debris to disperse.

In Fig.~\ref{fig:KI_disr_PCC}, we lay out the possible disruption configurations with launch dates spanning mid-2028 to a few months before 2024 YR4's lunar encounter in late 2032. Three mission trajectory cases, covering early, middle, and late launch dates, are presented, specifically in regions of Fig.~\ref{fig:KI_disr_PCC} that show the largest disruptable NEO mass. While launching earlier is preferable if possible, it is necessary to have backup launch options available in the event of a missed launch or failure to disrupt the asteroid. 

\begin{figure}[H]
\centering
\begin{subfigure}{0.48\textwidth}
    \centering
    \includegraphics[width=\textwidth]{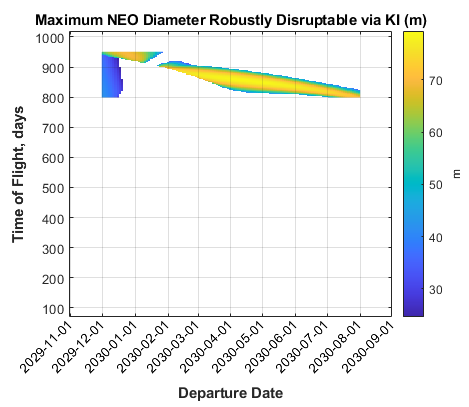}
    \caption{Case 1 (Early): PCC plot showing potential launch dates from December 2029 to August 2030 for KI disruption of 2024 YR$_{\text{4}}$}
    \label{fig:MaxNEOdisrupt_tof_case1}
\end{subfigure}
\hfill
\begin{subfigure}{0.48\textwidth}
    \centering
    \includegraphics[width=\textwidth]{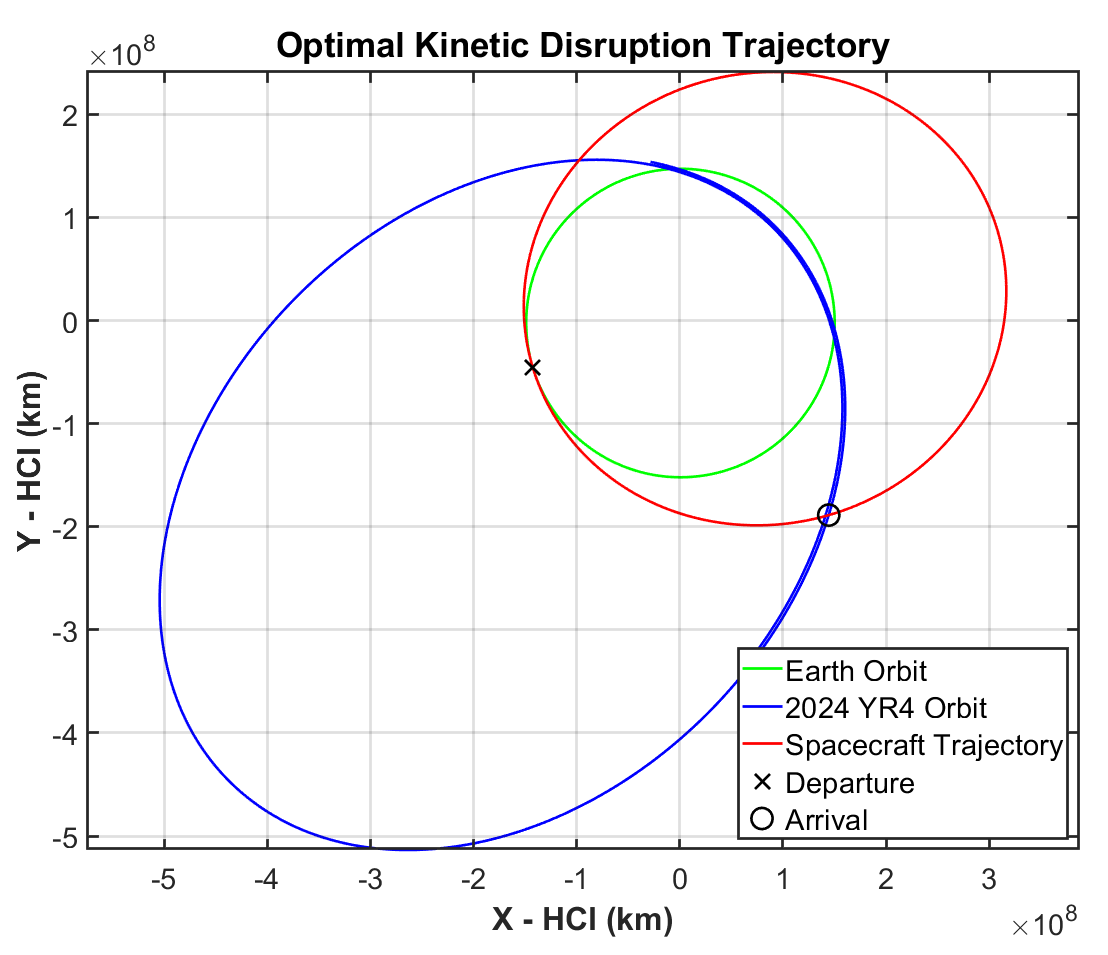}
    \caption{Exemplar 2024 YR$_{\text{4}}$ kinetic impactor trajectory launching April 2030 \phantom{this text will be invisible}}
    \label{fig:MaxNEOdisrupt_traj_case1}
\end{subfigure}
\caption{Case 1: PCC plot of kinetic impactor trajectories for an early (2030) launch to 2024 YR$_{\text{4}}$ and optimal trajectory from that set that maximizes the largest disruptable NEO}
\label{fig:disruption_case1}
\end{figure}

Case 1 includes a range of dates between December 2029 and August 2030, and assumes $\beta = 2$ and $\rho = 3.259$ g/cm$^3$. Fig.~\ref{fig:MaxNEOdisrupt_tof_case1} displays the most optimal trajectories that maximize the disruptable NEO diameter. For Case 1, the largest disruptable NEO is 78.5 meters, the launch date is 2030-04-08, and the arrival date of the impactor is 2032-08-13. Case 1 represents an early launch that would be able to disrupt up to the 99.7\% High Mass realization of the asteroid. The trajectory to 2024 YR$_{\text{4}}$ for this configuration of launch and arrival dates is shown in Fig.~\ref{fig:MaxNEOdisrupt_traj_case1}, and the column labeled ``Early'' in Table~\ref{tab:Case_results_summary} presents the trajectory parameters.

\begin{figure}[H]
\centering
\begin{subfigure}{0.48\textwidth}
    \centering
    \includegraphics[width=\textwidth]{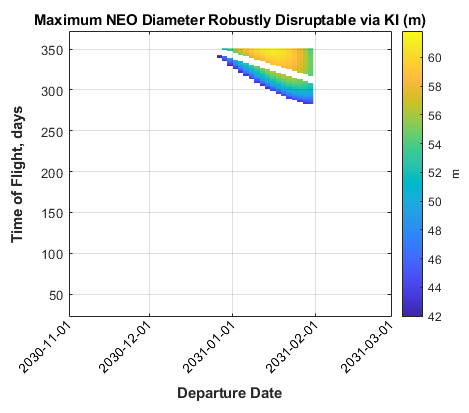}
    \caption{Case 2 (Middle): PCC plot showing potential launch dates from late Dec 2030 to late Jan 2031 for KI disruption of 2024 YR$_{\text{4}}$}
    \label{fig:MaxNEOdisrupt_tof_case2}
\end{subfigure}
\hfill
\begin{subfigure}{0.48\textwidth}
    \centering
    \includegraphics[width=\textwidth]{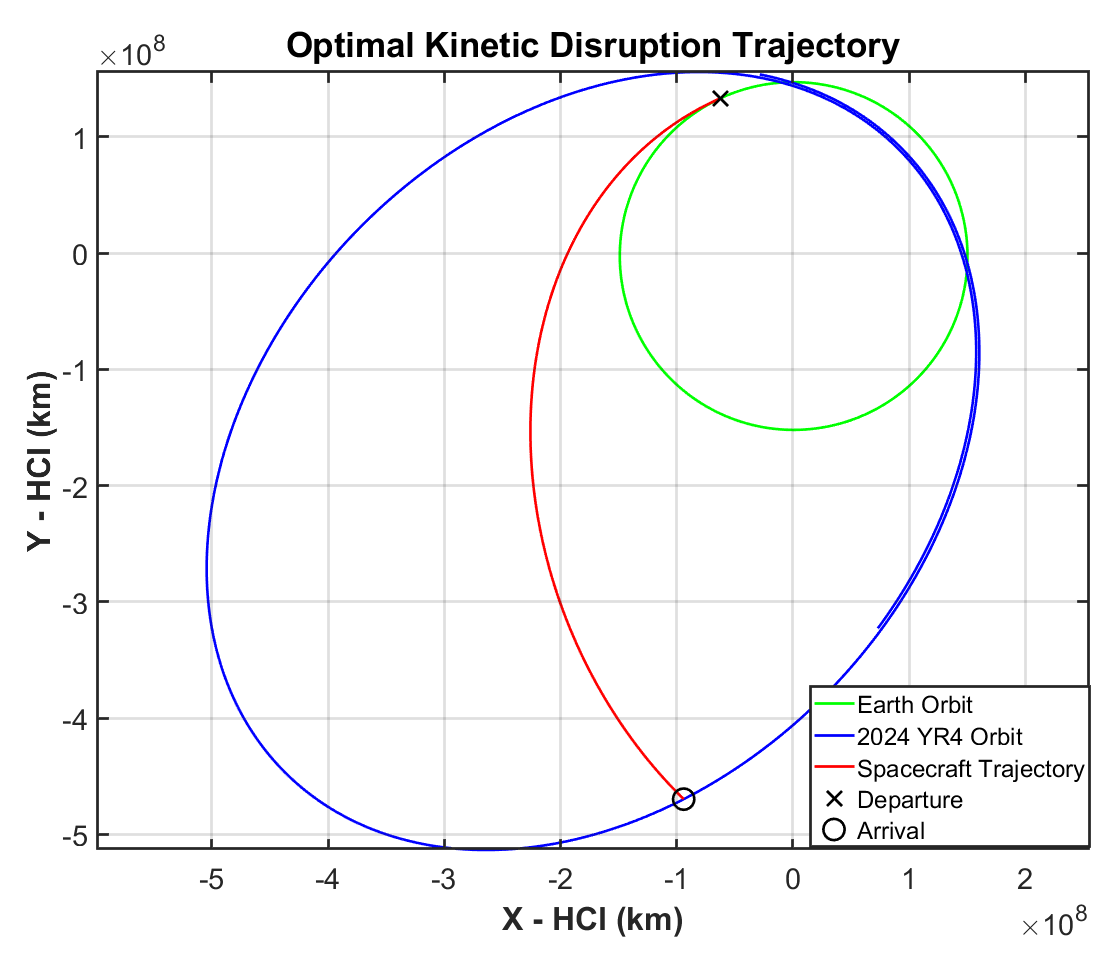}
    \caption{Exemplar 2024 YR$_{\text{4}}$ kinetic impactor trajectory launching January 2031 \phantom{this text will be invisible forever and ever and ever and ever and ever and ever}}
    \label{fig:MaxNEOdisrupt_traj_case2}
\end{subfigure}
\caption{Case 2: PCC plot of kinetic impactor trajectories for a middle (2031) launch to 2024 YR$_{\text{4}}$ and optimal trajectory from that set that maximizes the largest disruptable NEO}
\label{fig:disruption_case2}
\end{figure}

Case 2 targets a range of dates between November 2030 and the end of January 2031, where we set $\beta = 2$ and $\rho = 2.848$ g/cm$^3$. The departure date and time of flight configurations are shown in Fig.~\ref{fig:MaxNEOdisrupt_tof_case2} and the results of the most optimal configuration are listed in column ``Middle'' of Table~\ref{tab:Case_results_summary}. For Case 2, the largest disruptable NEO diameter is 61.2 meters, indicating that the kinetic impactor would be able to disrupt up to the 75\% High Mass realization of 2024 YR$_4$. The launch date of this mission profile would be 2031-01-16 and the arrival date is 2032-01-01. If a reconnaissance mission takes place before this launch date, it could determine whether the asteroid's mass is at or below the 75th percentile. If it is larger, this configuration of launch and arrival dates will not be feasible for robustly disrupting 2024 YR$_4$.

\begin{figure}[H]
\centering
\begin{subfigure}{0.48\textwidth}
    \centering
    \includegraphics[width=\linewidth]{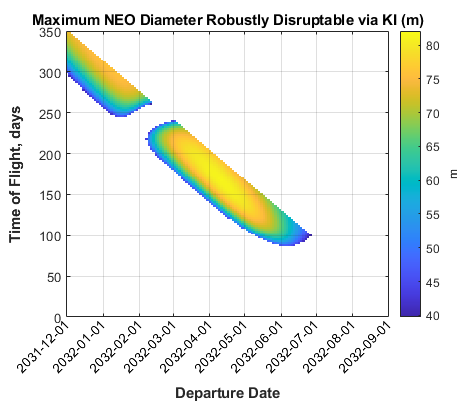}
    \caption{Case 3 (Late): PCC plot showing potential launch dates from Dec 2031 to Sept 2032 for KI disruption of 2024 YR$_{\text{4}}$}
    \label{fig:MaxNEOdisrupt_tof_case3}
\end{subfigure}
\hfill
\begin{subfigure}{0.48\textwidth}
    \centering
    \includegraphics[width=\linewidth]{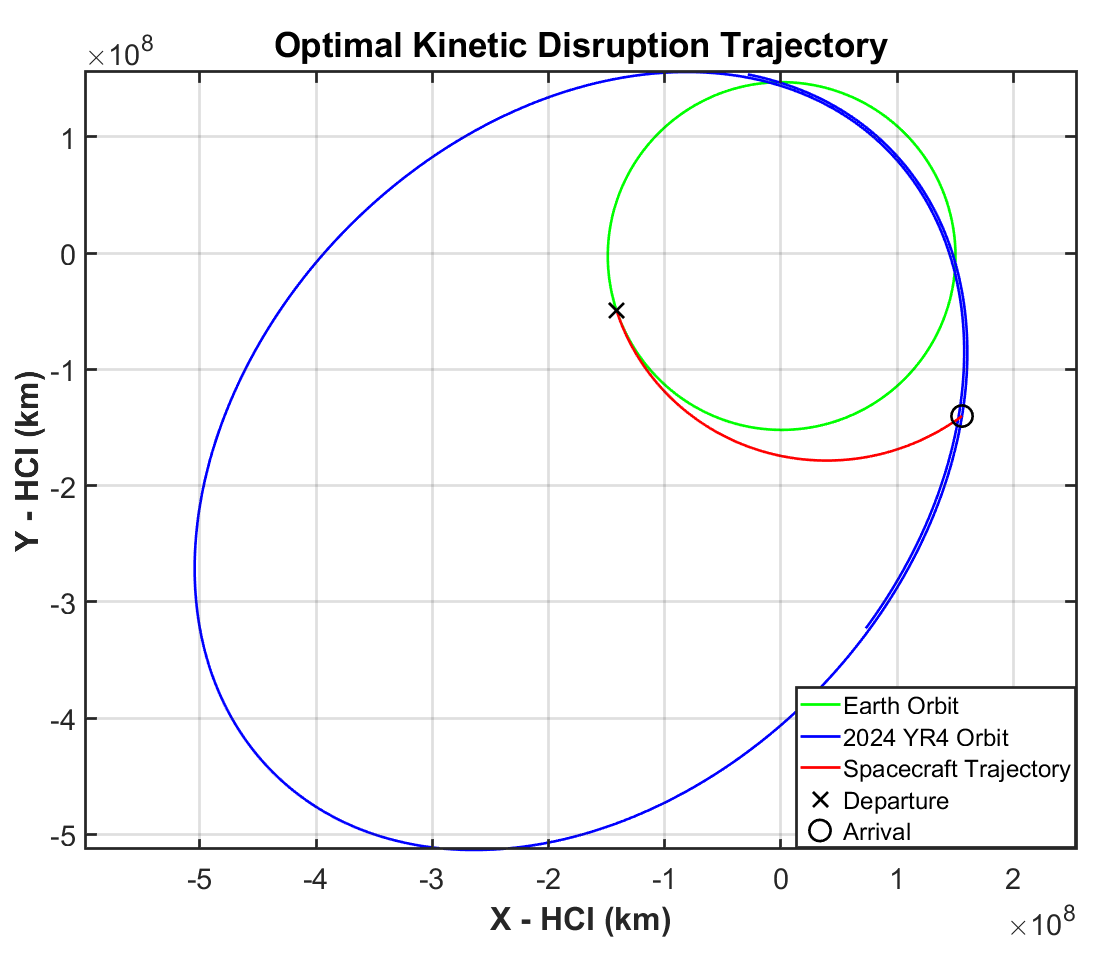}
    \caption{Exemplar 2024 YR$_{\text{4}}$ kinetic impactor trajectory launching April 2032 \phantom{this text will be invisible forever and ever and ever and ever, and ever and ever and ever and ever and ever and i guess we still need more space here omg}}
    \label{fig:MaxNEOdisrupt_traj_case3}
\end{subfigure}
\caption{Case 3: PCC plot of kinetic impactor trajectories for a late (2032) launch to 2024 YR$_{\text{4}}$ and optimal trajectory from that set that maximizes the largest disruptable NEO}
\label{fig:disruption_case3}
\end{figure}

Case 3 includes a range of dates between December 2031 and September 2032, three months before the lunar encounter, where $\beta = 2$ and $\rho = 3.259$ g/cm$^3$. Fig.~\ref{fig:MaxNEOdisrupt_tof_case3} displays the departure date and time of flight configurations for these latest possible launch times. Results for the optimal trajectory are summarized in the ``Late'' column of Table~\ref{tab:Case_results_summary}, and the optimal trajectory is displayed in Fig.~\ref{fig:MaxNEOdisrupt_tof_case3}. For a launch date of 2032-04-09 and arrival date of 2032-09-02, the maximum disruptable NEO diameter is 81.2 meters. This indicates that the kinetic impactor would be able to robustly disrupt up to at least the 99.7\% High Mass realization of 2024 YR$_4$, and potentially higher, according to Fig.~\ref{fig:KI_disr_PCC}

\begin{table}[htbp]
\centering
\caption{Combined Results Summary: Kinetic Disruption Cases 1--3}
\label{tab:Case_results_summary}
\begin{tabular}{lcccc}
\toprule
\text{} & \text{Early} & \text{Middle} & \text{Late} \\
\midrule
Best launch date        & 2030-04-08 & 2031-01-16 & 2032-04-09 \\
TOF (days)     & 858 & 350 & 146 \\
Best arrival date       & 2032-08-13 & 2032-01-01 & 2032-09-02\\
Launch $C_{3}$ (km$^{2}$/s$^{2}$)          & 26.38 & 60.82 & 12.13 \\
DLA                    & -7.63$^\circ$ & -12.57$^\circ$ & 5.35$^\circ$ \\
Arrival $V_{\infty}$ (km/s)    & 24.671 & 16.468 & 21.210\\
$D_{max}$ (m)    & 78.5 & 61.2 & 81.2\\
KI mass (kg)      & 8877 & 4012 & 11839 \\
SES angle               & 146.04$^\circ$ & 16.11$^\circ$ & 119.44$^\circ$\\
Approach pahse angle   & 3.42$^\circ$ & 2.43$^\circ$ & 6.95$^\circ$\\
Trajectory classification & 1 rev, short-way & 0 revs, short-way & 0 revs, short-way\\
Min distance to Sun (au) & 1.00 & 0.98 & 1.00\\
Max distance to Sun (au) & 2.14 & 3.20 & 1.40\\
\bottomrule
\end{tabular}
\end{table}

\subsection{Nuclear Robust Disruption Missions}

\begin{figure}
    \centering
    \includegraphics[width=0.7\textwidth]{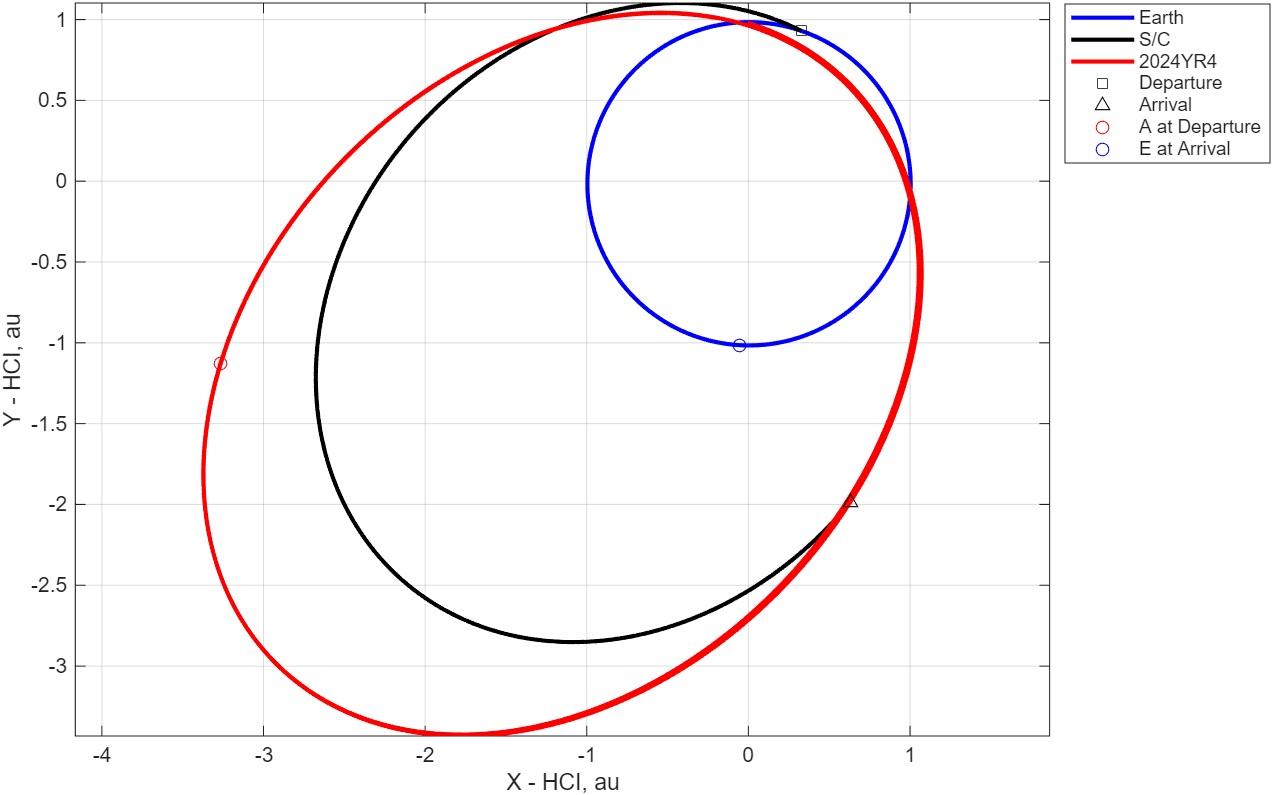}
    \captionof{figure}{Early launch option trajectory for nuclear disruption intercept mission to 2024 YR$_{\text{4}}$.}
    \label{fig:NED_disr_traj1}
\end{figure}

\begin{figure}
    \centering
    \includegraphics[width=0.7\textwidth]{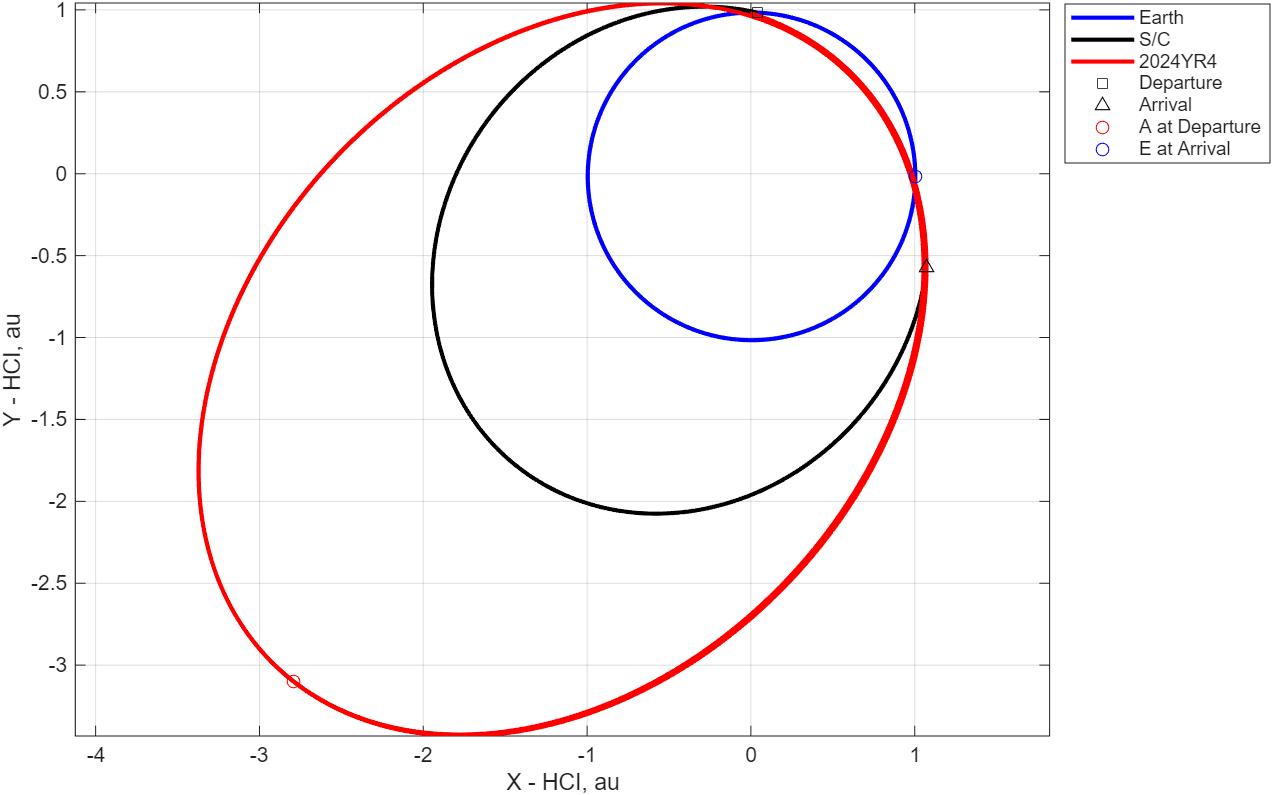}
    \captionof{figure}{Middle launch option trajectory for nuclear disruption intercept mission to 2024 YR$_{\text{4}}$.}
    \label{fig:NED_disr_traj2}
\end{figure}

\begin{figure}
    \centering
    \includegraphics[width=0.7\textwidth]{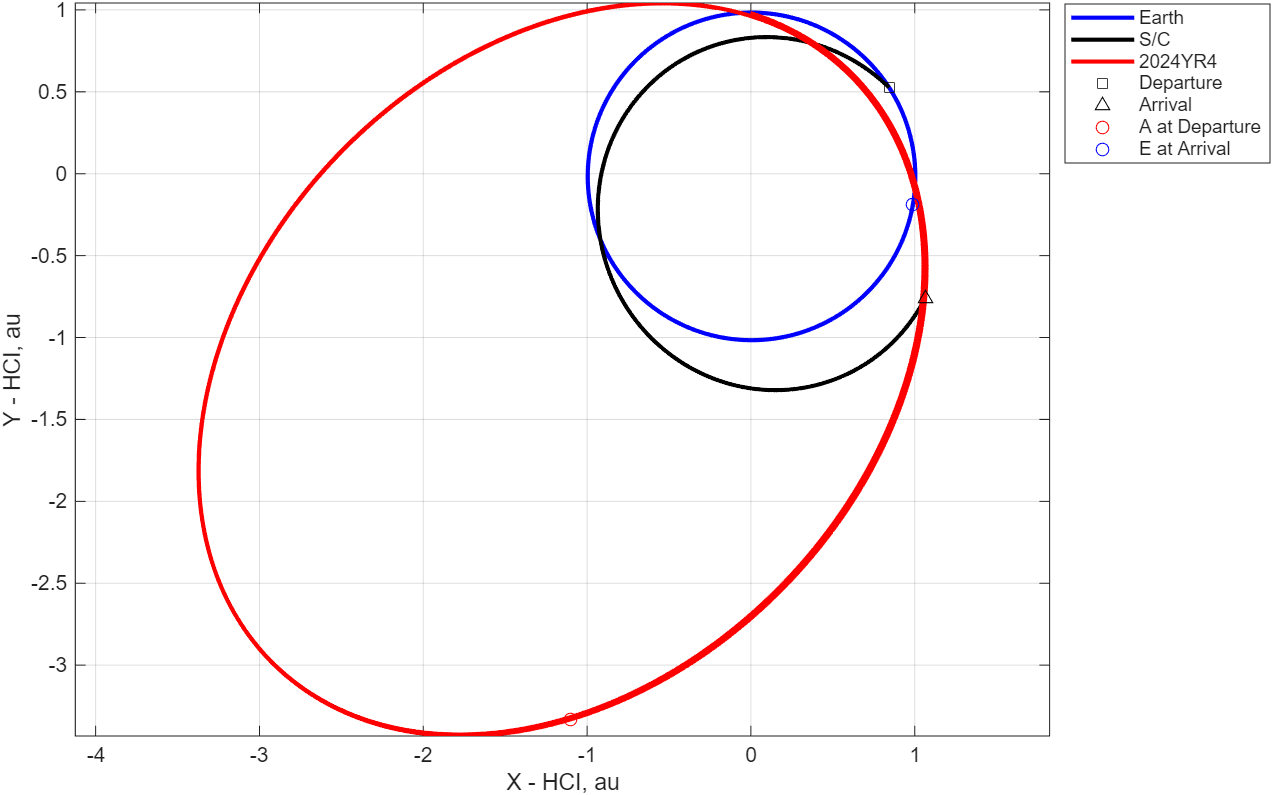}
    \captionof{figure}{Late launch option trajectory for nuclear disruption intercept mission to 2024 YR$_{\text{4}}$.}
    \label{fig:NED_disr_traj3}
\end{figure}

The 2024 YR$_{\text{4}}$ intercept trajectory trade spaces depicted in Figures~\ref{fig:PCCBallAll} and~\ref{fig:ChemFlybyParetoFront} were explored to identify several representative mission trajectory solutions for nuclear robust disruption via standoff detonation during high velocity terminal approach via radar fuzing. Based on the radar fuzing analysis in Section~\ref{ss:radar_fuze}, we set an upper limit of 15 km/s on the spacecraft's velocity relative to the asteroid at arrival. We also assume that at least 3000 kg of total spacecraft mass is required to deliver a 1 Mt NED to the asteroid. This is a rough order of magnitude mass estimate that can be refined in future work beyond the scope of this paper. Example trajectories, the parameters for which are listed in Table~\ref{tab:NED_disr_trajs}, were selected to represent early, middle, and late launch opportunities. Asteroid arrival speed was minimized to the extent possible. Figures~\ref{fig:NED_disr_traj1},~\ref{fig:NED_disr_traj2}, and~\ref{fig:NED_disr_traj3}  show the trajectories for the early, middle, and late launch options, respectively.

\begin{table}[htbp]
\centering
\caption{Exemplar Nuclear Disruption Mission Intercept Trajectories}
\label{tab:NED_disr_trajs}
\begin{tabular}{lccc}
\toprule
 & Early & Middle & Late \\
\midrule
Launch date                     & 2029-12-03 & 2030-12-20 & 2031-10-26 \\
TOF (days)                      & 928 & 642 & 322 \\
Arrival date                    & 2032-6-18 & 2032-09-22 & 2032-09-12 \\
Launch $C_3$ (km$^{2}$/s$^{2}$) & 70.749 & 69.067 & 50.426 \\
DLA                             & 12.47$^{\circ}$ & 6.21$^{\circ}$ & -23.46$^{\circ}$ \\
Max. Launch Mass (kg)           & 3019 & 3180 & 5232 \\
Arrival $V_\infty$ (km/s)       & 3.1 & 5.0 & 13.44 \\
Arrival SES angle               & 141.72$^{\circ}$ & 97.86$^{\circ}$ & 107.96$^{\circ}$ \\
Approach phase angle            & 12.23$^{\circ}$ & 14.77$^{\circ}$ & 19.88$^{\circ}$ \\
Trajectory type                 & 0 rev, long-way & 0 rev, long-way & 0 rev, long-way \\
Min. solar distance (au)        & 0.99 & 0.98 & 0.81 \\
Max. solar distance (au)        & 2.08 & 1.22 & 1.31 \\
Earth distance (au)             & 1.19 & 0.56 & 0.58 \\
Earth light-time (mins)         & 9.89 & 4.69 & 4.84 \\
\bottomrule
\end{tabular}
\end{table}

As with the prevously reported flyby analyses for OSIRIS-APEX and the purpose-built flyby reconnaissance spacecraft, in our modeling of the visual magnitude of the asteroid from the spacecraft's perspective during the approach trajectory, we use the H-G Slope formulation~\cite{dymock2007h}. Our modeling for the kinetic and nuclear robust disruption missions indicates that the DRACO instrument would detect the 99.7\% HPDI Low Mass realization of 2024 YR$_{\text{4}}$ (the dimmest realization of the asteroid) approximately 25 hours before intercept for the fastest approach trajectory, which is the Early kinetic robust disruption option (24.671 km/s approach speed). Detection occurred at much earlier times for the other kinetic and nuclear robust disruption mission trajectories, which have slower approach speeds. For terminal phase course corrections, target detection at least 12 hours before intercept is required. Handling later detection times may be possible in certain circumstances, but autonomous operations might be required. Further analysis would be needed to determine whether a kinetic or nuclear robust disruption mission employing DRACO or a DRACO-like terminal guidance sensor could truly perform a successful intercept of 2024 YR$_{\text{4}}$, but this early assessment indicates potential feasibility.

\section{Exemplar Mission Campaigns}\label{s:ex_missions}

Based on the trajectory options detailed in Section \ref{sec_trajOptions}, we lay out three candidate mission campaigns to characterize 2024 YR$_{\text{4}}$ and mitigate a possible lunar impact. These campaigns are not the only possible campaigns, but they represent the trade space of options while balancing feasibility of implementation. These campaigns propose purpose-built spacecraft. Although retasking an existing spacecraft appears plausible for reconnaissance in some cases, the uncertainties associated with their performance for missions they were not designed for makes them poorer candidates in this scenario. Additionally, we only consider robust disruption for impact prevention missions. As discussed above, the current uncertainties in 2024 YR$_{\text{4}}$'s physical properties make it difficult to design a KI for safe deflection, and those uncertainties will not be reduced in time to inform the design of the KI spacecraft. Meanwhile, there are no trajectories compatible with nuclear deflection that would give enough time to develop the spacecraft.

Example mission campaign option 1 consists of a flyby reconnaissance spacecraft and a KI robust disruption. The development of both spacecraft would begin in Q4 of CY2025, before the lunar impact is confirmed or ruled out. The flyby reconnaissance spacecraft would have a three year development time, launch in December 2028, and arrive in June 2029. The KI disruption spacecraft would launch in April 2030 after a $\sim$4.5 year development time, with the opportunity to make adjustments during the final year of development based on the data returned from the flyby mission. The KI disruption would occur in August 2032, $\sim$4 months before the lunar encounter. There is a backup KI launch opportunity in April 2032 if needed.

Example mission campaign option 2 augments campaign 1 with a rendezvous reconnaissance spacecraft to further characterize the asteroid and observe the KI disruption to confirm success and monitor post-disruption asteroid debris. The rendezvous spacecraft would begin development at the same time as the flyby spacecraft and launch in November 2029, giving $\sim$4 years of development time. It would arrive in June 2032, $\sim$2 years after the KI spacecraft launches and $\sim$2 months before the KI spacecraft disrupts the asteroid.

Example mission campaign option 3 forgoes the flyby spacecraft, and replaces the KI disruption mission with a nuclear disruption mission instead. This decreases the total number of spacecraft from campaign option 2, and potentially allows the decision to begin developing any spacecraft to be delayed until after the JWST attempted observation in early 2026. The rendezvous reconnaissance spacecraft for campaign option 3 is the same as campaign option 2 (November 2029 launch, June 2032 arrival). The nuclear disruption spacecraft would launch in December 2030 with a backup launch opportunity in October 2031. It would disrupt the asteroid in September 2032, $\sim$3 months before the lunar encounter. These campaigns are summarized in Table~\ref{tab:exemplar_mission_summary}.

Finally, we note that the flyby reconnaissance mission, which launches in December 2028 and arrives at the asteroid in June of 2029, could be flown as a rendezvous reconnaissance mission instead, because the speed relative to the asteroid at arrival is low enough ($\sim$0.951 km/s). While this would significantly improve reconnaissance performance and provide better information about the asteroid for use in planning the disruption mission, building a rendezvous-capable spacecraft in time for a December 2028 launch could prove too challenging, even if it had Q4 CY2025 start.

\begin{table}[htbp]
\centering
\caption{Summary of Exemplar Mission Campaign Options}
\label{tab:exemplar_mission_summary}
\begin{tabular}{llcccc}
\toprule
             &        & Flyby    & Rendezvous & Kinetic    & Nuclear  \\
             &        & Reconnaissance    & Reconnaissance     & Disruption & Disruption  \\
\midrule
\multirow{2}{*}{Campaign Option \#1} & Launch & 2028-Dec & \multirow{2}{*}{X}          & 2030-Apr   &  \multirow{2}{*}{X} \\
             & Arrive & 2029-Jun &            & 2032-Aug   &   \\
\midrule
\multirow{2}{*}{Campaign Option \#2} & Launch & 2028-Dec & 2029-Nov   & 2030-Apr   &  \multirow{2}{*}{X} \\
             & Arrive & 2029-Jun & 2032-Jun   & 2032-Aug   &   \\
\midrule
\multirow{2}{*}{Campaign Option \#3} & Launch &  \multirow{2}{*}{X}        & 2029-Nov   &  \multirow{2}{*}{X}          & 2030-Dec \\
             & Arrive &          & 2032-Jun   &            & 2032-Sep  \\
\bottomrule
\end{tabular}
\end{table}

\section{Conclusion}\label{s:conc}

In this work, we have provided an overview of deflection and robust disruption mechanics and techniques, with an emphasis on impulsive techniques utilizing Kinetic Impactors (KIs) or standoff detonation of Nuclear Explosive Devices (NEDs). We have described the notional requirements for deflection or robust disruption of 2024 YR$_{\text{4}}$, based on current knowledge and accounting for current uncertainties in asteroid physical properties (principally mass), lunar impact location (should the asteroid indeed be on a lunar-impacting trajectory), and deflection/disruption system performance (e.g., the momentum enhancement factor, $\beta$, for KIs). We have analyzed and discussed spacecraft mission trajectory options for flyby reconnaissance, rendezvous reconnaissance, deflection, and robust disruption. We have also generated exemplar mission campaign options based on the trajectory analysis results, noting key dates and discussing when associated decisions would need to be made in order for missions to be deployed. The best reconnaissance mission options launch in late 2028, leaving only approximately three years for development at the time of this writing in August 2025. Deflection missions were assessed and appear impractical. However, kinetic robust disruption missions are available with launches between April 2030 and April 2032. Nuclear robust disruption missions are also available with launches between late 2029 and late 2031.

When considering the various mission options we describe herein, it is important to keep in mind that 2024 YR$_{\text{4}}$'s lunar impact probability currently stands at $\sim$4\%. A JWST detection of the asteroid is currently planned for February of 2026~\cite{jwst_proposal}. If that JWST observation is successful, the resulting reduction of uncertainties in the asteroid's orbit would likely either significantly reduce or increase the lunar impact probability, but not rule out or confirm lunar impact. If that observation is not successful, then it would not be until sometime around June of 2028, when ground-based telescopes can detect the asteroid again, that the 2032 lunar impact probability would be refined. The lunar impact would most likely be ruled out or in during the 2028 observations, but it is also possible that the impact probability would remain greater than 0 or less than 1. Any decisions about sending missions to 2024 YR$_{\text{4}}$ made prior to June 2028 would need to be made in the face of significant uncertainty regarding whether the lunar impact will actually happen. A successful JWST detection of the asteroid in early 2026 would reduce but not eliminate that uncertainty. However, the JWST detection will be challenging and success is not guaranteed. Waiting to make mission deployment decisions until 100\% certainty about lunar impact is obtained around or after June 2028 will eliminate some mission options that require earlier commitment.


Thought may be given to conditional mission development plans that incorporate off-ramps and/or feature re-purposing options, such that development of 2024 YR$_{\text{4}}$ missions could begin prior to June 2028 without resulting in wasted resources if the lunar impact is ruled out. We also note that a 2024 YR$_{\text{4}}$ reconnaissance mission could be deployed even if lunar impact is ruled out. This could be the 2028 flyby recon launch shown in Campaign Options 1 and 2, or it could be the rendezvous reconnaissance mission from Campaign Option 3 in Table~\ref{tab:exemplar_mission_summary}, which could be operated as a flyby mission instead. A launch in November 2029 would provide roughly 11 additional months for spacecraft development compared to the earlier 2028 flyby recon launch, while still allowing the spacecraft to reach 2024 YR$_{\text{4}}$ about 6 months before its lunar encounter, and at a low relative speed. This would enable valuable data collection on the asteroid while also satisfying the requirement for a rapid-response demonstration. Deploying a reconnaissance mission to the asteroid even if lunar impact is ruled out would be for the purposes of a) satisfying the recommendation in the chapter on planetary defense in the National Academies of Science, Engineering, and Medicine’s Decadal Strategy for Planetary Science and Astrobiology 2023-2032~\cite{OWL_2023} stating that ``The highest priority planetary defense demonstration mission to follow DART and NEO Surveyor should be a rapid-response, flyby reconnaissance mission targeted to a challenging NEO, representative of the population ($\sim$50--100 m in diameter) ...'' and b) making significant progress towards actions assigned to NASA and other agencies for rapid response NEO reconnaissance capability development in the White House’s National Preparedness Strategy \& Action Plan for Near-Earth Object Hazards and Planetary Defense~\cite{Natl_PD_Plan_2023} (see Actions 3.1 and 3.2), and reiterated in the NASA Planetary Defense Strategy and Action Plan~\cite{NASA_PD_Plan_2023} (see Actions 3.1 and 3.2).

%
%

\subsection{Utilization of Extant Spacecraft for Reconnaissance}

When assessing the feasibility of utilizing extant spacecraft for reconnaissance of 2024 YR$_{\text{4}}$, we concentrated on several U.S. spacecraft: Janus, OSIRIS-APEX, Psyche, and Lucy (for which we found re-tasking to 2024 YR$_{\text{4}}$ isn't feasible). 

The Janus spacecraft---currently partially disassembled and in storage after its originally planned rideshare launch was delayed for reasons beyond the Janus team's control---could potentially be utilized for reconnaissance of 2024 YR$_{\text{4}}$. Trajectories for flyby reconnaissance of 2024 YR$_{\text{4}}$ that do not exceed the Janus spacecraft's solar distance limits of $\sim$1.0 to 1.6 au are available with launch during the first half of June 2028, or within a narrower period during the first week of December 2028. Additional launch opportunities are available in 2032 that respect Janus's solar distance limits but involve higher flyby speeds than the 2028 opportunities. Those launch dates are in early to mid 2032 and lead to encountering the asteroid just a few months before the potential lunar impact date. However, there would be ample time to construct purpose-built reconnaissance spacecraft for the 2032 launch opportunities, meaning they shouldn't be considered Janus-specific opportunities. Finally, additional analyses would be needed to determine whether the Janus spacecraft instrument payload and GNC system would be suitable for 2024 YR$_{\text{4}}$ flyby reconnaissance, particularly given the relatively high flyby speeds involved and the small size of 2024 YR$_{\text{4}}$. The amount of lead time necessary to reassemble the Janus spacecraft and prepare it for launch would also need to be determined.

The OSIRIS-APEX spacecraft could be diverted sometime between September 2026 and June 2027 for a flyby of 2024 YR$_{\text{4}}$ in late October or early November 2028, with flyby speed between 8 and 11 km/s. Diverting OSIRIS-APEX in this way would make it unable to maintain its current prime mission to rendezvous with Apophis in June 2029. However, if diverted on September 7, 2026 for the 2024 YR$_{\text{4}}$ flyby in 2028, it would then be possible for OSIRIS-APEX to perform an $\sim$13 km/s flyby of Apophis on March 26, 2029 (a few weeks before Apophis's historic Earth close approach on April 13, 2029). Or, if diverted on November 17, 2026, then after the 2028 flyby of 2024 YR$_{\text{4}}$ OSIRIS-APEX could go on to perform an $\sim$11 km/s flyby of Apophis on September 18, 2030. The OSIRIS-APEX spacecraft is designed for rendezvous rather than high-speed flyby; further work would therefore be needed to assess the feasibility of using it for flyby reconnaissance.

The Psyche spacecraft could be diverted in June 2026 for a rendezvous with 2024 YR$_{\text{4}}$ in December 2030, or diverted in July 2026 for a 8.7 km/s flyby of 2024 YR$_{\text{4}}$ in January 2029. Either of these options would mean sacrificing the prime mission of rendezvous with the main belt asteroid Psyche. The Psyche spacecraft is designed for rendezvous rather than high-speed flyby; further work would therefore be needed to assess the feasibility of using it for flyby reconnaissance.




\subsection{Other Types of Deflection Systems}


The small size of 2024 YR$_{\text{4}}$ puts it in a regime where Ion Beam Deflection (IBD) might in principle be an effective approach, and it should be considered. However, the lack of suitable rendezvous trajectory opportunities, due to the high eccentricity of the asteroid's orbit, combined with the very short time available for spacecraft development, cruise, and deflection operations, makes IBD a less favorable deflection approach for this object. Nevertheless, as described earlier, there are rendezvous trajectories launching in later 2028 that can deliver a moderately sized spacecraft to a rendezvous with 2024 YR$_{\text{4}}$ in late 2029. Such a spacecraft could carry a $\sim$20 kW solar array and have hundreds of kilograms of extra xenon propellant that could be used for deflection. While deflection produced by such a spacecraft at over 3.3 au from the Sun would be far from optimal, it nevertheless could amount to many hundreds of kilometers of deflection in the Earth's B-plane before the propellant ran out (performing analysis to confirm this is future work). Since a distance of many hundreds of kilometers is a significant fraction of the entire lunar impact risk chord, such an IBD mission could be sufficient to avoid the impact, depending on the lunar impact location and associated deflection distance requirement, in which case IBD could be a viable deflection option. Note that a Gravity Tractor (GT) spacecraft would provide less deflection than an IBD spacecraft, all else being equal.





\subsection{Future Work}

\begin{itemize}
    \item Assessment of what additional mission options may be enabled by $C_3 > 100$ km$^2$/s$^2$ launch capabilities, e.g., an extra propulsion stage being provided for a spacecraft launching on a Falcon Heavy Expendable.
    \item Assessment of the Guidance, Navigation, and Control (GNC) requirements, including sensors and actuators, for a) successful execution of a flyby reconnoiter for 2024 YR$_{\text{4}}$, and b) successful execution of a KI or NED delivery mission that precisely targets 2024 YR$_{\text{4}}$ at high approach speeds.
    \item Assessment of build times for the various spacecraft called for by the mission options considered herein. The potential lunar impact occurring in 2032, only $\sim$7 years from now, combined with 2024 YR$_{\text{4}}$'s $\sim$4-year orbit period, severely limit the time available for building spacecraft to be deployed to the asteroid, especially considering that typical spacecraft build times (ATP to launch) for interplanetary missions are $\sim$3--5 years.
    \item Assessment of the ability to utilize reconnaissance mission data about the asteroid to inform the design and construction of lunar impact prevention missions. The compressed timeline in this scenario severely limits opportunities to gather and utilize reconnaissance data to inform the design of deflection or disruption missions.
    \item Further development of deflection/disruption mission design strategies to overcome uncertainties in asteroid mass, lunar impact location (i.e. deflection distance), and achieved $\Delta V$ on the asteroid (manifesting as uncertainty in $\beta$ for KIs and an analogous uncertainty in achieved $\Delta V$ for standoff NED detonations). This is of particular utility in case of 2024 YR$_{\text{4}}$ because of the aforementioned limitations in the time available to acquire and utilize reconnaissance data.
\end{itemize}

\bmhead{Acknowledgements}
Part of this research was conducted at the Jet Propulsion Laboratory, California Institute of Technology, under a contract with the National Aeronautics and Space Administration (80NM0018D0004).

Portions of this work were performed under the auspices of the U.S. Department of Energy by Lawrence Livermore National Laboratory under Contract DE-AC52-07NA27344. LLNL-JRNL-2010771

This research was carried out in part at Sandia National Laboratories. Sandia National Laboratories is a multi-mission laboratory managed and operated by National Technology \& Engineering Solutions of Sandia, LLC (NTESS), a wholly owned subsidiary of Honeywell International Inc., for the U.S. Department of Energy’s National Nuclear Security Administration (DOE/NNSA) under contract DE-NA0003525. This written work is authored by an employee of NTESS. The employee, not NTESS, owns the right, title and interest in and to the written work and is responsible for its contents. Any subjective views or opinions that might be expressed in the written work do not necessarily represent the views of the U.S. Government. The publisher acknowledges that the U.S. Government retains a non-exclusive, paid-up, irrevocable, world-wide license to publish or reproduce the published form of this written work or allow others to do so, for U.S. Government purposes. The DOE will provide public access to results of federally sponsored research in accordance with the DOE Public Access Plan.

%
%
%
%
%


\bibliography{references}

\begin{thebibliography}{10}
\providecommand{\doi}[1]{\url{https://doi.org/#1}}
\bibcommenthead

\bibitem[\protect\citeauthoryear{{Farnocchia, D. et al.}}{2025}]{Farnocchia_2025}
{Farnocchia, D  et al }.
\newblock The impact hazard assessment for near-Earth asteroid 2024 YR4.
\newblock The Journal of the Astronautical Sciences. 2025;.

\bibitem[\protect\citeauthoryear{Wiegert et~al.}{2025}]{Wiegert_2025}
Wiegert P, Brown P, Lopes J, Connors M.: {The Potential Danger to Satellites due to Ejecta from a 2032 Lunar Impact by Asteroid 2024 YR4}.
\newblock Available from: \url{https://arxiv.org/abs/2506.11217}.

\bibitem[\protect\citeauthoryear{Rivkin et~al.}{2025}]{Rivkin_2025}
Rivkin AS, Mueller T, MacLennan E, Holler B, Burdanov A, de~Wit J, et~al.
\newblock {JWST Observations of Potentially Hazardous Asteroid 2024 YR4}.
\newblock Research Notes of the AAS. 2025 apr;9(4):70.
\newblock \doi{10.3847/2515-5172/adc6f0}.

\bibitem[\protect\citeauthoryear{Bolin et~al.}{2025}]{Bolin_2025}
Bolin BT, Hanuš J, Denneau L, Bonamico R, Abron LM, Delbo M, et~al.
\newblock {The Discovery and Characterization of Earth-crossing Asteroid 2024 YR4}.
\newblock The Astrophysical Journal Letters. 2025 apr;984(1):L25.
\newblock \doi{10.3847/2041-8213/adc910}.

\bibitem[\protect\citeauthoryear{{Dotson} et~al.}{2024}]{Dotson2024}
{Dotson} JL, {Wheeler} L, {Mathias} D.
\newblock {Consequences of asteroid characterization on the state of knowledge about inferred physical parameters and impact risk}.
\newblock Acta Astronautica. 2024 Sep;222:550--555.
\newblock \doi{10.1016/j.actaastro.2024.04.020}.

\bibitem[\protect\citeauthoryear{Barbee et~al.}{2018}]{Barbee_2018}
Barbee BW, Syal MB, Dearborn D, Gisler G, Greenaugh K, Howley KM, et~al.
\newblock Options and uncertainties in planetary defense: Mission planning and vehicle design for flexible response.
\newblock Acta Astronautica. 2018;143:37--61.
\newblock \doi{https://doi.org/10.1016/j.actaastro.2017.10.021}.

\bibitem[\protect\citeauthoryear{Dearborn et~al.}{2020}]{Dearborn_2020}
Dearborn DSP, {Bruck Syal} M, Barbee BW, Gisler G, Greenaugh K, Howley KM, et~al.
\newblock Options and uncertainties in planetary defense: Impulse-dependent response and the physical properties of asteroids.
\newblock Acta Astronautica. 2020;166:290--305.
\newblock \doi{https://doi.org/10.1016/j.actaastro.2019.10.026}.

\bibitem[\protect\citeauthoryear{Kumamoto et~al.}{2025}]{Kumamoto_2025}
Kumamoto KM, Barbee BW, Pearl JM, Syal MB.: Probing disruption heuristics for kinetic deflection of asteroids.
\newblock Poster presented at the 9th IAA Planetary Defense Conference, Stellenbosch, Cape Town, South Africa, May 5-9, 2025, \url{https://iaa.4hdt.ro/event/1/contributions/43/attachments/35/572/IAA-PDC-25-06-77-poster.pdf}.

\bibitem[\protect\citeauthoryear{{Holsapple} and {Housen}}{2019}]{Holsapple_2019}
{Holsapple} KA, {Housen} KR.
\newblock {The Catastrophic Disruptions of Asteroids: History, Features, New Constraints and Interpretations}.
\newblock Planetary and Space Science. 2019 Dec;179:104724.
\newblock \doi{10.1016/j.pss.2019.104724}.

\bibitem[\protect\citeauthoryear{{Stewart} and {Leinhardt}}{2009}]{Stewart_2009}
{Stewart} ST, {Leinhardt} ZM.
\newblock {Velocity-Dependent Catastrophic Disruption Criteria for Planetesimals}.
\newblock ApJL. 2009 Feb;691(2):L133--L137.
\newblock \doi{10.1088/0004-637X/691/2/L133}.

\bibitem[\protect\citeauthoryear{Farnocchia}{2015}]{Farnocchia_2015}
Farnocchia D.
\newblock Impact hazard monitoring: theory and implementation.
\newblock Proceedings of the International Astronomical Union. 2015;10(S318):221–230.
\newblock \doi{10.1017/S1743921315007310}.

\bibitem[\protect\citeauthoryear{{Chodas}}{1999}]{Chodas1999}
{Chodas} PW.
\newblock {Orbit uncertainties, keyholes, and collision probabilities.}
\newblock In: Bulletin of the American Astronomical Society. vol.~31; 1999. p. 1117.

\bibitem[\protect\citeauthoryear{{Cheng, A. F., Agrusa, H. F., Barbee, B. W. et al.}}{2023}]{Cheng_2023}
{Cheng, A  F , Agrusa, H  F , Barbee, B  W  et al }.
\newblock Momentum transfer from the DART mission kinetic impact on asteroid Dimorphos.
\newblock Nature. 2023;616:457–460.
\newblock \doi{https://doi.org/10.1038/s41586-023-05878-z}.

\bibitem[\protect\citeauthoryear{King et~al.}{2021}]{King2021LateTimeDisruptions}
King PK, Bruck~Syal M, Dearborn DSP, Managan R, Owen JM, Raskin C.
\newblock Late-time small body disruptions for planetary defense.
\newblock Acta Astronautica. 2021;188:367--386.
\newblock \doi{10.1016/j.actaastro.2021.07.034}.

\bibitem[\protect\citeauthoryear{Managan and Burkey}{2025}]{Managan_2025}
Managan RA, Burkey MT.: Revisions to the High Fluence NEO Deflection Formulae.
\newblock Poster presented at the 9th IAA Planetary Defense Conference, Stellenbosch, Cape Town, South Africa, May 5-9, 2025, \url{https://iaa.4hdt.ro/event/1/contributions/48/attachments/38/362/IAA-PDC-25-06-45-poster.pdf}.

\bibitem[\protect\citeauthoryear{Ulaby et~al.}{2019}]{Ulaby_2019}
Ulaby FT, Dobson MC, A\'lvarez-Pe\'rez JL.
\newblock Handbook of radar scattering statistics for terrain.
\newblock Boston: Artech House; 2019.

\bibitem[\protect\citeauthoryear{Bull et~al.}{2025}]{Bull_2025}
Bull RA, Vavrina MA, Lyzhoft J, Atchison J, Barbee B.
\newblock Retasking In-flight Spacecraft for Rapid Response Reconnaissance in Planetary Defense Exercises.
\newblock In: Planetary Defense Conference; 2025. .

\bibitem[\protect\citeauthoryear{Shoer et~al.}{2022}]{Shoer_2022}
Shoer J, Hopkins JB, Bierhaus EB, Brack D, McCaa TC, Waldorff K, et~al.
\newblock Janus: Launch of a NASA SmallSat Mission to Near-Earth Binary Asteroids.
\newblock In: Small Sat Conference; 2022. .

\bibitem[\protect\citeauthoryear{Scheeres et~al.}{2024}]{scheeres2024}
Scheeres D, Bierhaus B, May L.
\newblock Potential Future Missions for NASA's Janus Spacecraft.
\newblock 45th COSPAR Scientific Assembly Held 13-21 July. 2024;45:243.

\bibitem[\protect\citeauthoryear{Chabot et~al.}{2024}]{Chabot2024}
Chabot NL, Atchison JA, Bull R, Rivkin AS, Daly RT, Ballouz RL, et~al.
\newblock A Mission to Demonstrate Rapid-Response Flyby Reconnaissance for Planetary Defense.
\newblock In: International Astronautical Congress. IAC-24-E10; 2024. .

\bibitem[\protect\citeauthoryear{Lauretta et~al.}{2017}]{Lauretta2017}
Lauretta D, Balram-Knutson S, Beshore E, Boynton W, Drouet~d’Aubigny C, DellaGiustina D, et~al.
\newblock OSIRIS-REx: sample return from asteroid (101955) Bennu.
\newblock Space Science Reviews. 2017;212:925--984.

\bibitem[\protect\citeauthoryear{DellaGiustina et~al.}{2023}]{Dellagiustina2023}
DellaGiustina DN, Nolan MC, Polit AT, Moreau MC, Golish DR, Simon AA, et~al.
\newblock OSIRIS-APEX: an OSIRIS-REx extended mission to asteroid apophis.
\newblock The Planetary Science Journal. 2023;4(10):198.

\bibitem[\protect\citeauthoryear{Nolan et~al.}{2024}]{nolan2024osiris}
Nolan MC, DellaGiustina DN, Golish DR, Moreau MC, Polit AT, Simon AA, et~al.: {The OSIRIS-APEX Mission to Apophis}.
\newblock Lunar and Planetary Laboratory, University of Arizona (1629 E. University Blvd, Tucson, AZ 85721, USA) and NASA Goddard Space Flight Center (Greenbelt, MD, 20771, USA).
\newblock Contact: mcn1@arizona.edu.

\bibitem[\protect\citeauthoryear{Mink}{2015}]{mission2015sawg}
Mink RG.: {Origins Spectral Interpretation Resource Identification Security-Regolith Explorer (OSIRIS-REx) Project Mission Requirements Document}.
\newblock OSIRIS-REx-RQMT-0001 Revision J April 14, 2015, \url{https://repository.arizona.edu/bitstream/handle/10150/647685/MRD_J_SAWG.pdf}.

\bibitem[\protect\citeauthoryear{Dymock}{2007}]{dymock2007h}
Dymock R.
\newblock {The H and G magnitude system for asteroids}.
\newblock Journal of the British Astronomical Association, vol 117, no 6, p 342-343. 2007;117:342--343.

\bibitem[\protect\citeauthoryear{Mainzer et~al.}{2023}]{Mainzer_2023}
Mainzer AK, Masiero JR, Abell PA, Bauer JM, Bottke W, Buratti BJ, et~al.
\newblock The Near-Earth Object Surveyor Mission.
\newblock The Planetary Science Journal. 2023 Dec;4(12):224.
\newblock \doi{10.3847/PSJ/ad0468}.

\bibitem[\protect\citeauthoryear{Ernst et~al.}{2022}]{DRACO_SIS_2022}
Ernst C, Daly T, Barnouin O, Espiritu R.
\newblock Didymos Reconnaissance and Asteroid Camera for {OpNav} ({DRACO}) Uncalibrated/Calibrated Data Product Software Interface Specification.
\newblock Johns Hopkins University Applied Physics Laboratory; 2022.
\newblock Document for the DART (Double Asteroid Redirection Test) mission, \url{https://pdssbn.astro.umd.edu/holdings/pds4-dart:document_draco-v1.0/jhuapl_dart_draco_uncalibrated_calibrated_sis_v1.pdf}.

\bibitem[\protect\citeauthoryear{{Rivkin, A. S., et al.}}{2025}]{jwst_proposal}
{Rivkin, A  S , et al }.: Enabling early decision making for a possible lunar impact in 2032.
\newblock JWST Proposal. Cycle 4, ID. \#9441, \url{https://www.stsci.edu/jwst-program-info/download/jwst/pdf/9441/}.

\bibitem[\protect\citeauthoryear{{National Academies of Sciences, Engineering, and Medicine}}{2023}]{OWL_2023}
{National Academies of Sciences, Engineering, and Medicine}.: {Origins, Worlds, and Life: A Decadal Strategy for Planetary Science and Astrobiology 2023-2032}.
\newblock Washington, DC: The National Academies Press, \url{https://doi.org/10.17226/26522}.

\bibitem[\protect\citeauthoryear{{White House Office of Science and Technology Policy}}{2023}]{Natl_PD_Plan_2023}
{White House Office of Science and Technology Policy}.: {National Preparedness Strategy and Action Plan for Near-Earth Object Hazards and Planetary Defense}.
\newblock Planetary Defense Interagency Working Group, April 2023, \url{https://assets.science.nasa.gov/content/dam/science/psd/planetary-science-division/2025/2023-NSTC-National-Preparedness-Strategy-and-Action-Plan-for-NearEarth-Object-Hazards-and-Planetary-Defense.pdf}.

\bibitem[\protect\citeauthoryear{{National Aeronautics and Space Administration}}{2023}]{NASA_PD_Plan_2023}
{National Aeronautics and Space Administration}.: {NASA Planetary Defense Strategy and Action Plan}.
\newblock April 2023, \url{https://www3.nasa.gov/sites/default/files/atoms/files/nasa_-_planetary_defense_strategy_-_final-508.pdf}.

\end{thebibliography}

\end{document}